\documentclass[graybox]{svmult}

\usepackage{amssymb}
\usepackage{graphicx}
\usepackage[T1]{fontenc}
\fontencoding{T1}  
\usepackage[utf8]{inputenc}
\usepackage{subfigure}

\usepackage{url}

\graphicspath{{figs/}}

\begin{document}

\mainmatter  

\title*{On buildings that compute. A proposal}

\titlerunning{On buildings that compute}
\authorrunning{Adamatzky, Szaci{\l}owski, Przyczyna, Konkoli, Sirakoulis, Werner}

\author{
Andrew Adamatzky
\and 
Konrad Szaci{\l}owski
\and
Dawid Przyczyna
\and 
Zoran Konkoli
\and 
Georgios Ch. Sirakoulis
\and 
Liss C. Werner
}

\institute{
Andrew Adamatzky
\at
Unconventional Computing Laboratory, UWE Bristol, UK
\and 
Konrad Szaci{\l}owski
\at AGH University of Science and Technology, Academic Centre for Materials and Nanotechnology, Krak\'ow, Poland
\and
Dawid Przyczyna
\at 
 AGH University of Science and Technology, Faculty of Physics and Applied Computer Science , Krak\'ow, Poland
 \at AGH University of Science and Technology, Academic Centre for Materials and Nanotechnology, Krak\'ow, Poland
\and 
Zoran Konkoli
\at
Chalmers University of Technology, Department of Microtechnology and Nanoscience, 
G\"{o}thenburg, Sweden
\and Georgios Ch. Sirakoulis
\at
Department of Electrical \& Computer Engineering,
Democritus University of Thrace, Xanthi, Greece
\and Liss C. Werner
\at 
Institute of Architecture, Technical University of Berlin, Germany
}

\maketitle

\abstract{
We present ideas aimed at bringing revolutionary changes on architectures and buildings of tomorrow by radically advancing the technology for the building material concrete and hence building components. We propose that by using nanotechnology we could embed computation and sensing directly into the material used for construction. Intelligent concrete blocks and panels advanced with stimuli-responsive smart paints are the core of the proposed architecture. In particular, the photo-responsive paint would sense the buildings internal and external environment while the nano-material-concrete composite material would be capable of sensing the building environment and implement massive-parallel information processing resulting in distributed decision making. A calibration of the proposed materials with in-materio suitable computational methods and corresponding building information modelling, computer-aided design and digital manufacturing tools could be achieved via models and prototypes of information processing at nano-level. The emergent technology sees a building as high-level massive-parallel computer --- assembled of computing concrete blocks. Based on the generic principles of neuromorphic computation and reservoir computing we envisage a single building or an urban quarter to turn into a large-scale sensing substrate. It could behave as a universal computer, collecting and processing environmental information \emph{in situ} enabling appropriate data fusion. The broad range of spatio-temporal effects include infrastructural and human mobility, energy, bio-diversity, digital activity, urban management, art and socializing, robustness with regard to damage and noise or real-time monitoring of environmental changes. The proposed intelligent architectures will increase sustainability and viability in digitised urban environments by decreasing information transfer bandwidth by e.g, utilising 5G networks. The emergence of socio-cultural effect will create a cybernetic relationship with our dwellings and cities.  
}

\section{Introduction}
\label{introduction}


Nowadays, trends of smart homes are entering in our everyday lives~\cite{chan2008review,alam2012review,augusto2006designing,maass2002real,brush2018smart}. Multinational enterprises tend to providing virtual integration not only in our smart phones but also in smart speakers installed in our houses. With the addition of home appliances participating in the Internet-of-things (embedded with electronics, software, sensors, actuators and connectivity) our houses are becoming smarter (more responsive to their inhabitants)~\cite{li2011smart,darianian2008smart}. Even before they are built we can --- with the assistants of virtual reality and augmented reality --- experience the spaces we are planning to inhabit in future. With the help of touch-sensitive virtual environments our senses start merging with the matter to be applied in the material world outside of the ubiquitous world of bits and bytes. 

What if the process could go either way and also the building could sense us? What if intelligent matter of our surrounding could understand us humans, give us feedback and communicate with us. What if the walls surrounding us were not only supporting our roofs, but had increased functionality, i.e. sensing, actuating, calculating, processing, communicating and producing power? What if each brick\footnote{We are well aware that modern building technologies have a wide range of building blocks and panels; however, we are using word `brick' here for simplicity and readability reasons}, or building block, was a self-powered, decentralised computing entity that would comprise a part of an emerging, large-scale parallel computation? Then our smart buildings would be transformed to intelligent, computing, cerebral organisations that we could not only live in but also interact in a holistic cybernetic way. Those organisations would also offer an active unparalleled protection from, e.g. crime or natural disasters, could warn us against a dangerous structural damage hidden deeply in the walls, or simply give us joy by sending a jolly image to our human visual apparatus. In this chapter we discuss how to build such homes. 

We address the challenge of implementation of new building materials in buildings that are capable of power production, sensory readings extraction, communication and computation. The vision is moving from concurrent smart homes that have limited sensing capabilities and can connect to the internet, to advanced computing homes of the future that have advanced capabilities in sensing, actuating, processing, communications and computing. This will be achieved by employing materials with embedded information processing abilities and multiple, elementary devices interconnected in a massive grid.

We envisage embedding sensing and associated cognitive abilities directly in a building material and implement computation at several levels of hierarchy. Similar to the human body each millimetre of a concrete block/brick will be able to sense its environment; sensory processing and decision making is done at the level of a single block/brick, more complex problems (shape recognition, learning, predictions) are solved by a cluster of blocks/bricks co-operating in a wall. Thus a building becomes a super-computer at macro-scale while walls will be massive-parallel array processors at a meso-scale and each component of a wall acts as massive-parallel computing device at micro-scale encompassing in-memory computing. 

\begin{figure}[!tbp]
    \centering
    \includegraphics[width=0.95\textwidth]{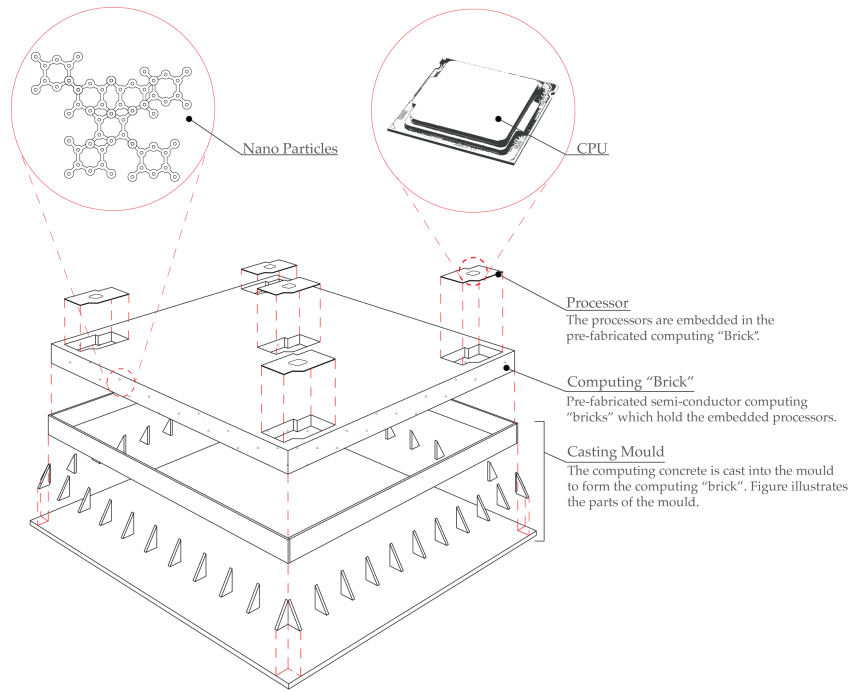}
    \caption{Prototyping computing architectures. Multi-scale computation implies generation of information at the micro-scale in the material, propagation and modulation of information by passing through higher structural elements: the concrete block, wall, building. See further descriptions in the text.}
    \label{fig:scheme}
\end{figure}

Key concepts of the computing architecture are illustrated in Fig.~\ref{fig:scheme}. Functional nanoparticles with photo-, chemo- and electro-sensitive and a range of electrical properties spanning all possible electronic elements are mixed in a concrete. Building blocks are made of the concrete. The building blocks are also equipped with processors for gathering information from distributed sensory elements, aids in decision making and location communication and enabling them to perform advanced computing coupled with modern in-memory principles. The blocks are assembled into a wall, which constitutes a massive parallel array processor.

These our  ideas are in tune Susan Stepney et al. thinking on gardening cyber-physical cities \cite{stepney2012gardening} however we adopted more pragmatic component of computing materials. 

The chapter is structured as follows. Section~\ref{computingconcrete} presents our suggestion to develop new semiconducting nano-materials capable of advanced sensing and information processing. In Sect.~\ref{computingbrick} we present the theoretical framework of massive-parallel computing and reservoir computing, the principle foundations of an efficient and effective computing system to be applied firstly on a wall-scale, and in a next step on a building scale. Section~\ref{computingwalls} suggests using modular standard sized blocks in order to establish the massively-parallel array processor and a communication grid between the blocks that are equipped with high-level processing units. In Section~\ref{computingarchitecture} we engage with an architectural vision featuring computing architectures and in Section~\ref{discussion} we discuss our overall vision as presented analytically before.

\section{Computing concrete}
\label{computingconcrete}

Novel semiconducting nanomaterials capable of advanced sensing and information processing include construction and decorative materials, in particular we focus on nanomaterials on titanium dioxide and other nanopigments already present in concrete and wall-paints featuring photo-catalytic, photoelectric and memristive properties. We will develop novel in-materio computing schemes, including fusion of binary, ternary and fuzzy logic~\cite{nanoscale, KS_ternary, KS_fuzzy, KS_JACS}, with recurrent neural networks~\cite{williams1989learning,mandic2001recurrent} and memristive synaptic devices~\cite{chang2011synaptic,kuzum2013synaptic,park2015electronic}. The dynamics of charge carriers within semiconducting nanostructures~\cite{ulbricht2011carrier}, combined with charge trapping properties of carbonaceous nanostructures~\cite{KS_neuro} will be utilised in construction of unique distributed reservoir computing systems with flexible connections between individual nodes. The large fraction of sensing and computing will be performed in specially designed materials incorporated into or deposited onto construction materials (concrete blocks, bricks, parge coat and wall paints). Material-based sensing will address light/temperature sensing, sounds/vibrations, and some pre-defined chemical species (e.g. carbon monoxide or other combustion products).

Wide band-gap semiconductors belong to the most widely studied materials. Whereas studies on photovoltaics, solar fuels and heterogeneous photocatalysis involving semiconducting nanoparticles seem to dominate current material science, the information-processing aspect of nanoscale semiconducting materials is still underrepresented.
Semiconducting nanomaterials offer a tremendous diversity and versatility of material properties, including optical and electrical properties, which can be utilised for computational purposes. Some semiconducting nanomaterials, including titanium dioxide, cadmium sulfide and lead iodide perovskites has been already employed in various computing architectures, including binary logic \cite{nanoscale} (either simple logic gates or more complex computing circuits, like reconfigurable logic gates~\cite{KS_JACS}, demultiplexers~\cite{KS_demult} or binary half adders)~\cite{KS_sum}, ternary logic~\cite{KS_ternary}, fuzzy logic~\cite{KS_fuzzy} and  neuromorphic computing~\cite{KS_neuro}. Moreover, the same class of materials, and with very similar surface modifications, have been used for advanced sensing applications~\cite{KS_sensors}. Furthermore, despite their photocatalytic activity, some organic-inorganic hybrid materials are exceptionally stable towards photo- and electrodegradation. Therefore we postulate, that this class of materials, fulfilling the paradigms of in-materio computing, is suitable as concrete/paint additives which will add computational power to construction materials.
These materials, however (like any other information-processing system) cannot operate without sources of data and energy. Due to photovoltaic properties of wide band gap semiconductors they can work in computing devices as the information processing structures and power generators at the same time. Furthermore, due to appropriate chemical structure of the surface, semiconducting nanostructures based on wide band gap oxide materials should easily integrate with classical construction materials, e.g. concrete. Thus, we should obtain hybrid construction materials capable of collecting information from users and environment, at the same time performing some forms of computation and yielding desired response to the users via their internal electrical activity. They should be able to monitor their own internal structure (e.g. detect local increase of humidity or cracks within the construction material). All these functions will be possible due to three facts:
\begin{itemize}
\item compatibility of inorganic wide band gap semiconductors with currently used construction materials such as concrete, clay bricks, steel
\item responsiveness of these materials to various stimuli, mainly optical and electrical, but also mechanical and chemical 
\item the ability of implementation of complex logic functionality including ternary and fuzzy logic on the basis of simple physical phenomena.
\end{itemize}

\begin{figure}[!tbp]
    \centering
    \subfigure[]{\includegraphics[width=0.49\textwidth]{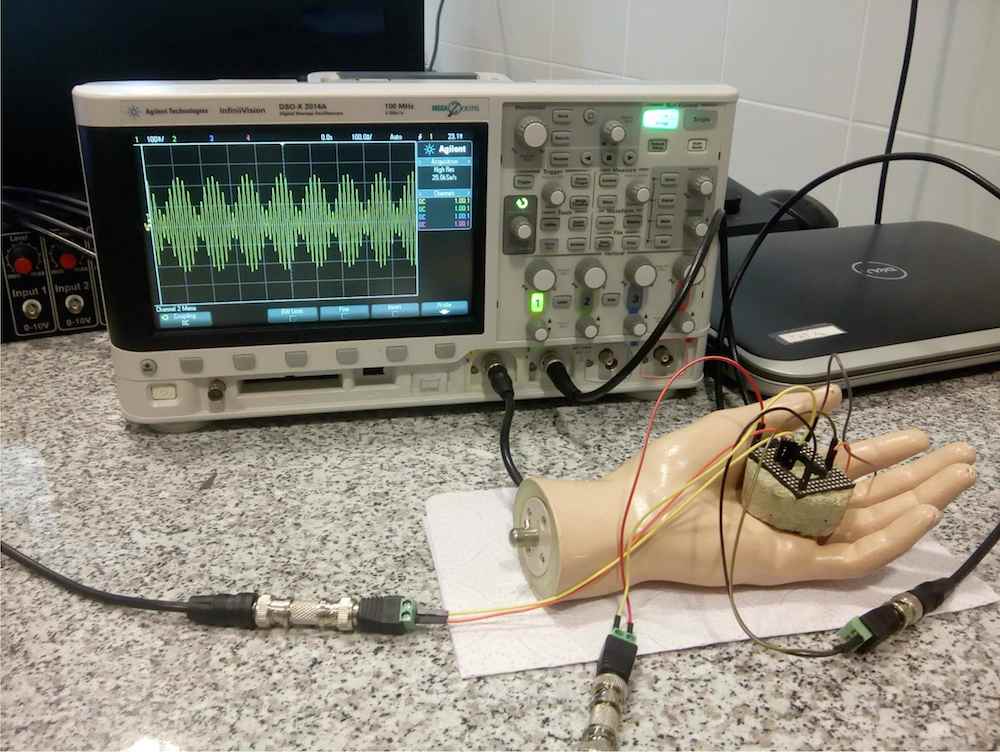}}
    \subfigure[]{\includegraphics[width=0.485\textwidth]{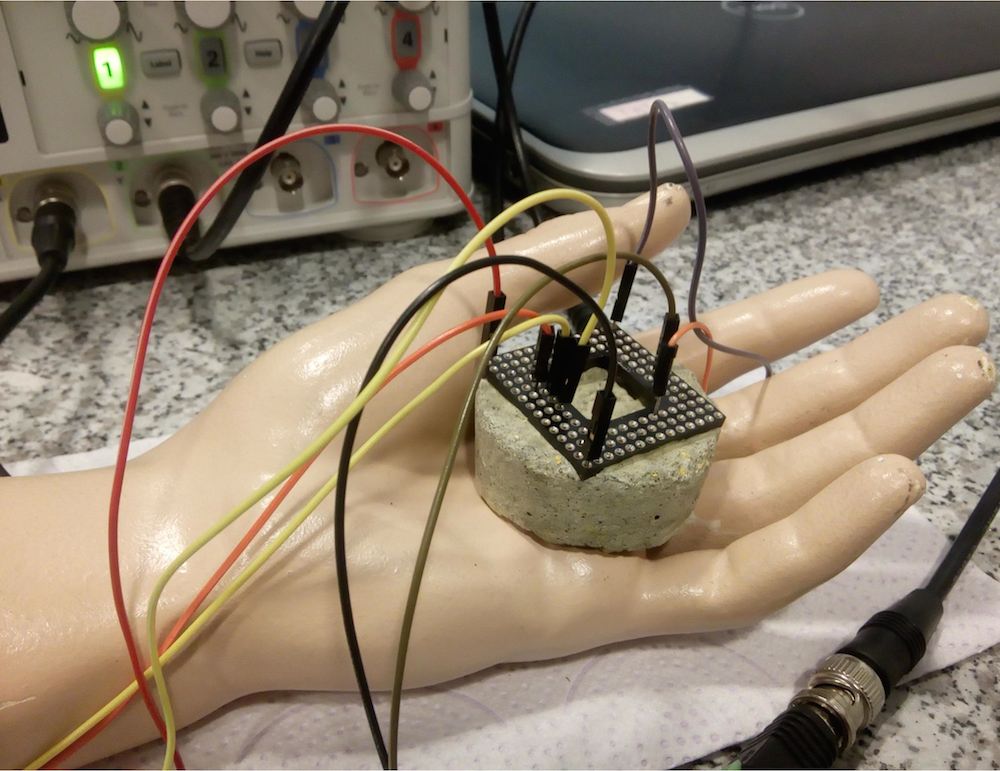}}
    \caption{A  photo taken during signal modulation using a piece of nanoparticle-doped concrete (a) and a close-up view of ‘concrete computing element' (b).}
    \label{fig:hand}
\end{figure}

Our preliminary study indicates, that traditional concrete-based materials can be used for advanced electrical signal processing. Concrete containing ca. 10\% percent of aluminium and steel shavings and 5\% of semiconducting nanoparticles --- antimony sulfoiodide nanowires carbon nanotubes and cadmium sulfide nanoparticles --- gains unique electrical properties, which can be approximated by distributed random resistor-capacitor network with some hysteresis. The latter results from slow ionic movements within the concrete matrix  combined with rapid electric response resulting from sub-percolation threshold arrangement of metallic particles embedded in concrete matrix, which may act as a complex signal conditioning circuit (Fig.~\ref{fig:hand}).

    \begin{figure}[!tbp]
    \centering
\subfigure[]{\includegraphics[width=0.45\textwidth]{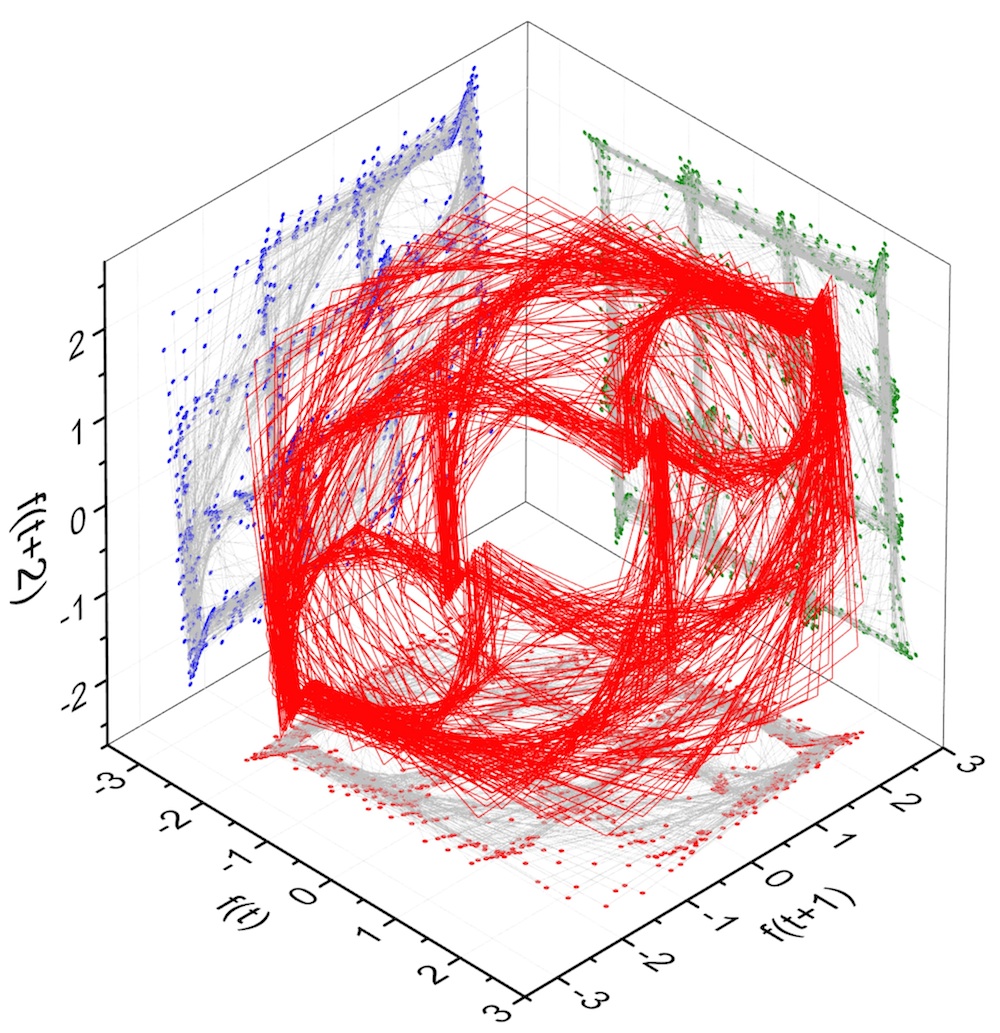}}
\subfigure[]{\includegraphics[width=0.45\textwidth]{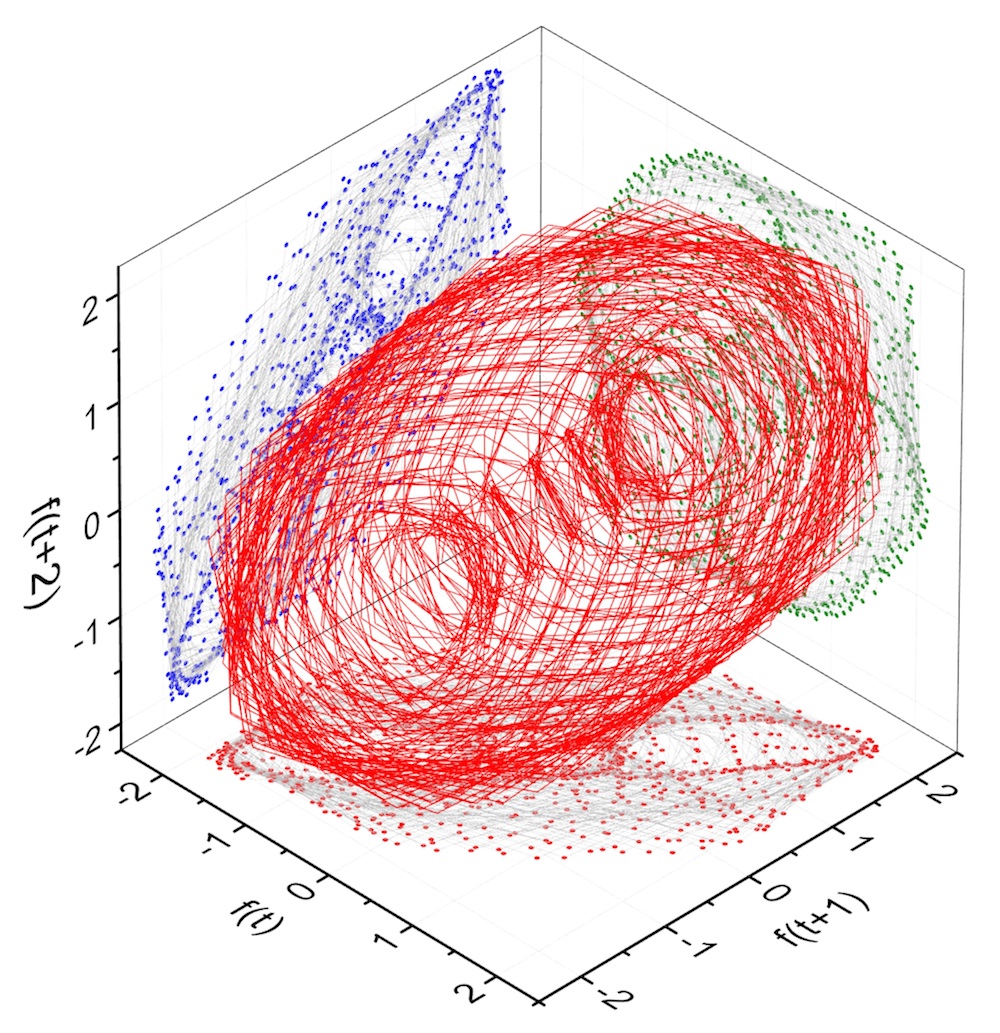}}
\subfigure[]{\includegraphics[width=0.45\textwidth]{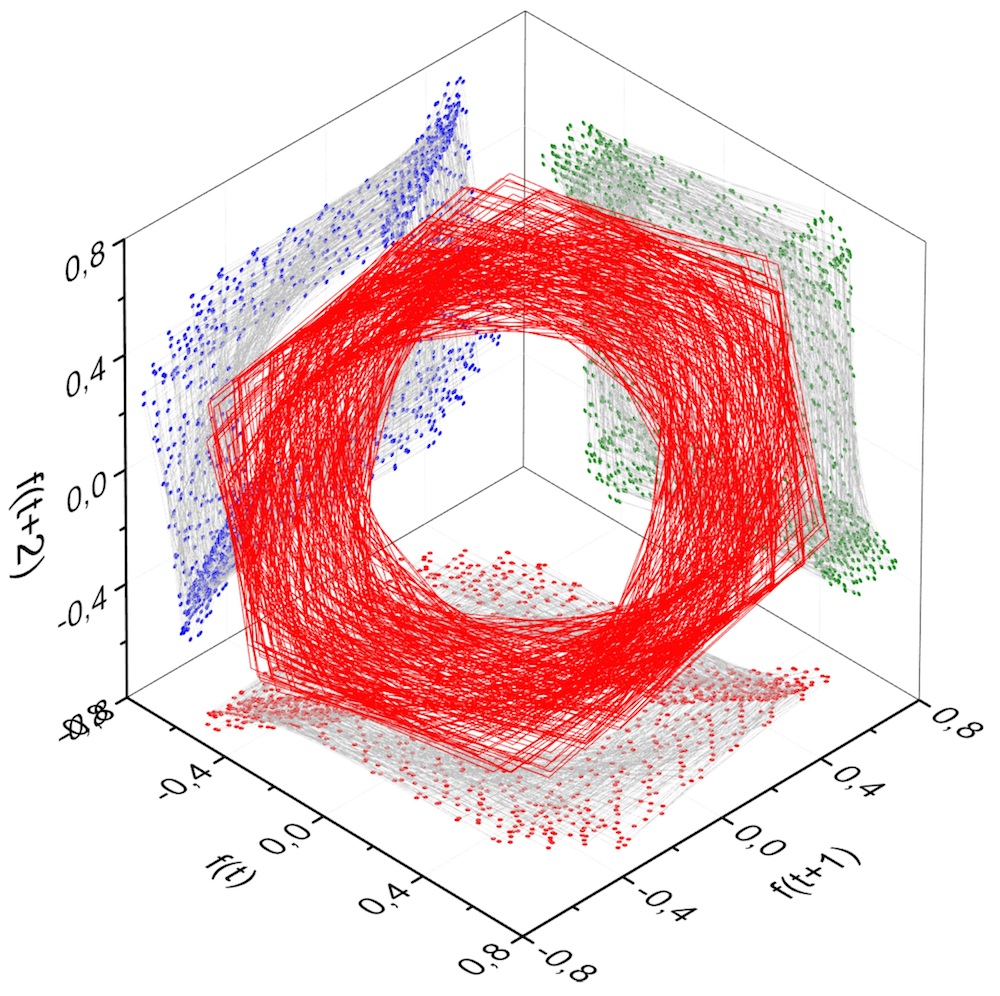}}
    \caption{Chaotic-like attractors produced by the `concrete-based microprocesor’ upon stimulation with the following signal combinations: (a)~square wave and square wave, (b)~square wave and sine wave, (c)~square wave and saw tooth (c). The same connectivity was used in all these cases.}
    \label{fig:attractor}
\end{figure}

The device was subjected to square wave, saw tooth and sine wave stimulation. Two independent arbitrary signal generators, connected with randomly chosen contacts, were used to stimulate the device. The square wave generator was tuned to the frequency of 100~Hz, and the other generator (square wave, sine wane, saw tooth) was tuned to the frequency of 101~Hz. An output signal was recorded also at randomly chosen pin of the pin-grid array socket.
The response of the system becomes extremely complex due to superposition of capacitive and resistive coupling within the concrete block and also due to an interplay of electrical and ionic conductivity, like in ion-diffusion memristors. The complexity of the response of ‘concrete-based microprocessor’ is shown in Fig.~\ref{fig:attractor}.

  \begin{figure}[!tbp]
    \centering
\subfigure[]{\includegraphics[width=0.59\textwidth]{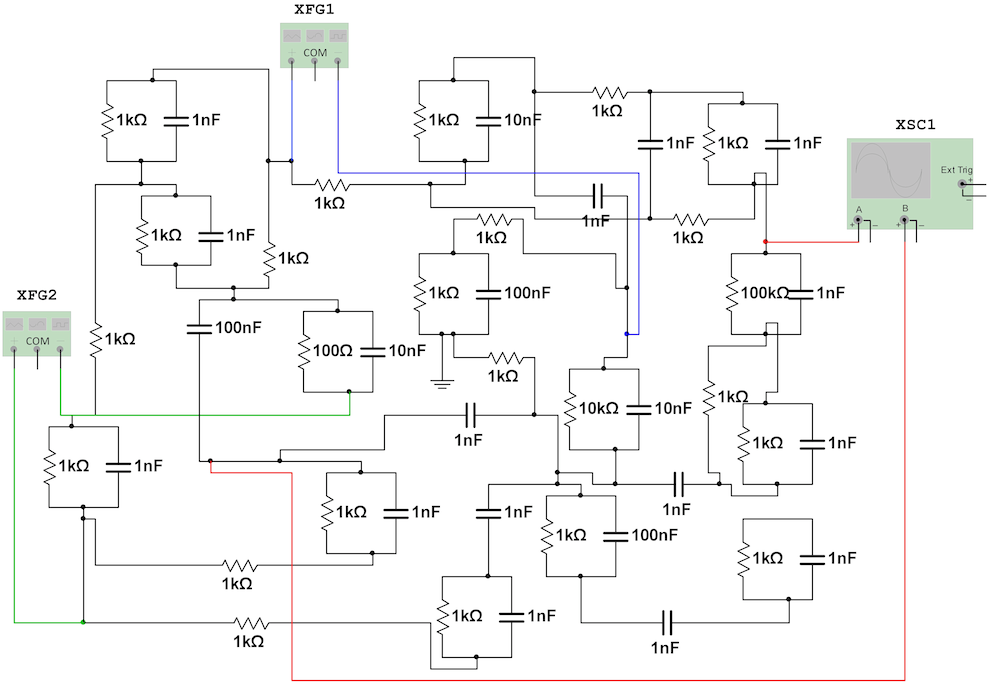}}
\subfigure[]{\includegraphics[width=0.39\textwidth]{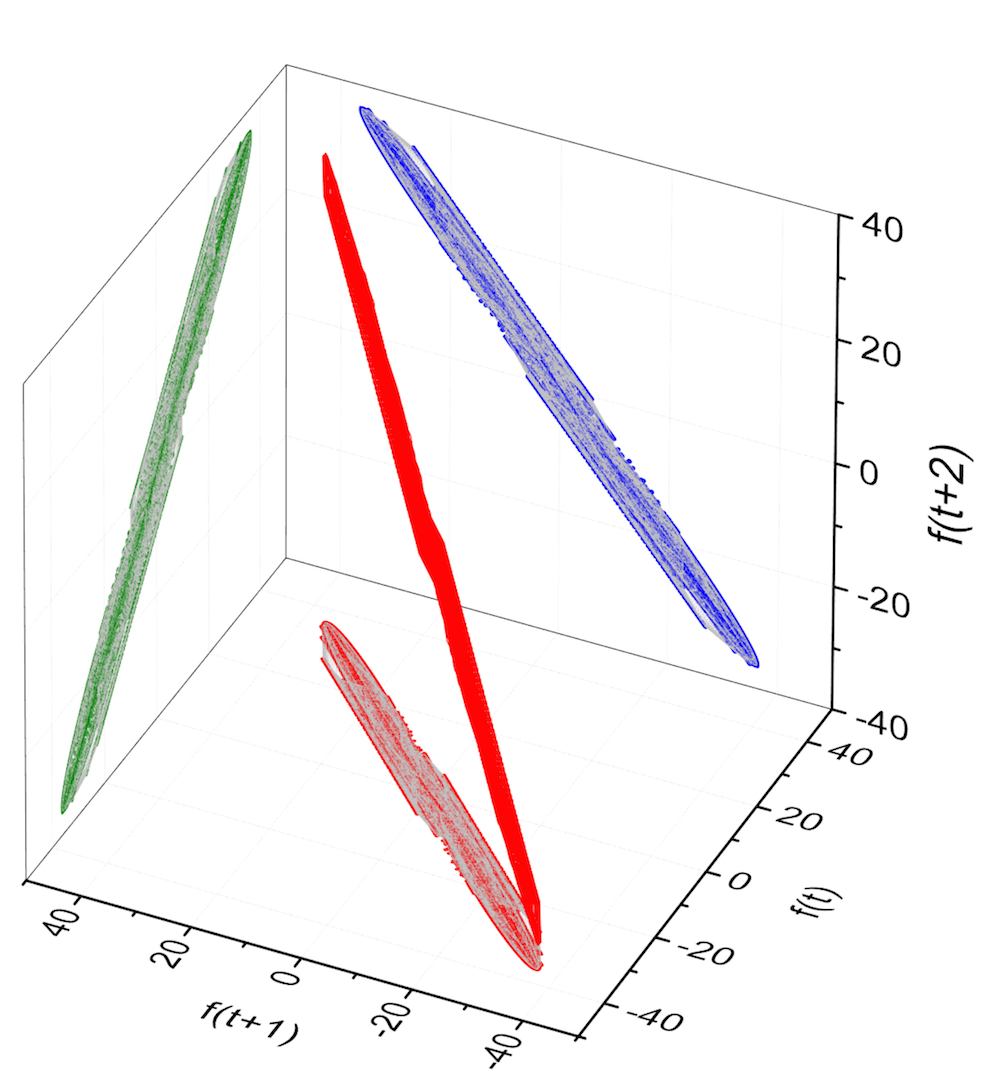}}
    \caption{A schematics of the random resistor–capacitor network takes as (a)~a model of metal- and semicondictor-doped 'computing concrete' and (b)~an attractor produced by the model upon stimulation with the square wave and sine wave combination.}
    \label{fig:circuit}
\end{figure}

The quasi-chaotic character is a result of nonlinear characteristic of some of the junctions present in the material, possibly of memristive character. A simple model, implemented in Multisim (Fig.~\ref{fig:circuit}) yields much less complex attractors.

This suggests that a concrete block with appropriate admixtures of metallic macro-particles and semiconducting nanoparticles offers significantly complex nonlinear behaviour  and  internal dynamics, associated with capacitive effects superimposed on some ion diffusion processes, that can be used as a computing node in a reservoir computing system. This dynamics, in turn, can be utilized for computation provided that the formal requirements of computation in physical systems are fulfilled~\cite{physical_system}. In the worst case novel hybrid materials will still provide distributed multisensory sensing system and will provide signal processing capabilities.  

Furthermore, in accordance to the aforementioned reconfigurability of the proposed nanomaterials, logic configurable circuits that constitute a key architecture for integrated nanoscale electronics can be developed in the  memristive networks \cite{Vourkas2014}. Such memristive logic circuits can overcome the disadvantages of memristive implication logic (stateful logic) \cite{Williams}, namely the necessity to perform lengthy sequences of stateful logic operations in order to synthesize a given Boolean function. The focus then should be firstly on hybrid transistor-memristor structures (together with logic gate implementation and/or the device-level requirements), where the logical state is represented as a voltage, instead of the normally used (mem)resistance, applying different logic circuit schemes like memristor-based combinational logic circuit design \cite{Vourkas2014} or Memristor-Ratioed Logic (MRL). Moreover, all memristive based logic/computational circuits, i.e. computation of Boolean functions in memristor-only circuits \cite{Papandroulidakis2014} as well as logic computation parallelism with memristors \cite{Papandroulidakis2017} would be examined as feasible promising in-memory computing nano-architectures. As an alternative novel analog computing concepts utilizing massively-parallel computing architectures, enabled by networks of memristors could be also equally considered \cite{Ariadne}. Having in mind semiconducting nanomaterials, as the ones earlier described in this section, with possibly different switching abilities, could result to better performance of the proposed computing medium and thus permit productive interfering in computations. In principle, such network-based computations \cite{adamatzky2013memristor} meaning massively parallel computations within array-like structures which accommodate networks of memristive components \cite{Vourkas2014plus} could be also another option for enabling the computation in the proposed novel semiconducting nanomaterials. Open issues to be discussed in this last option for appropriate computing, include among others, the network initialization, the requested number of devices to initialize and read, the suggested voltage supply, networks’ homogeneity and regularity and computation time and resulting devices densities \cite{Memristorbook}.

\section{Computing bricks}
\label{computingbrick}

During the design and prototyping of computing architectures we combine massive-parallel amorphous computing substrates (concrete block level) to act as an array processor with a regular grid-based structure (wall level) (Fig.~\ref{fig:scheme}). As an overarching computing principle, we propose to use reservoir computing. 

A typical reservoir computer features two parts, a dynamical system (the reservoir) that responds to an external signal (the input to the computation), and a readout layer that is used to analyse the state of the system (to produced the output of the computation)~\cite{maass2002real,lukovsevivcius2009reservoir,verstraeten2007experimental,lukovsevivcius2012reservoir}. The general idea is that the readout layer should be simple, i.e. the computation should be performed by the reservoir. The programming process involves tuning of the readout layer. This is usually done through a rather direct supervised learning procedure, that does not involve deep learning or elaborate back-propagation methods. Such artificial intelligence is easy to train, e.g., using gradient descent machine learning algorithms –-- towards a specific task, and does not require an extensive technological or engineering overhead to implement~\cite{RN697}. As a result,  reservoir computing has proven extremely useful for the design of a plethora of neuromorphic information processing applications that require fast real-time analysis of the time-series data~\cite{sillin2013theoretical,yi2016fpga,larger2017high,van2017advances,kasabov2016evolving}. 

Often, to build reservoir computing solutions, one simulates the operation of a reservoir computer on the standard digital computer. Interestingly, up to now there are very few practical implementation of reservoir computers in hardware. Up to date pioneering efforts exploit photonic systems~\cite{larger2012photonic,vandoorne2014experimental,vandoorne2008toward}, complex amorphous materials~\cite{dale2017reservoir,dale2016evolving,vissol2016data,burkow2016exploring,broersma2017computational}, etc. In here we suggest yet another computing substrate: the building and its components; walls, blocks or other pre-fabricated construction components, and ultimately the material it is made of. We suggest that innovative prototypical building materials offer new computing paradigms without advanced electronic engineering burden. There is no need to carefully micro-engineer each component individually, which would be hardly feasible, since the charge carrier dynamics in self-assembled structures automatically provides a molecular or nanoparticle-scale computing platform. While the sensing or computing performance of individual structure (chemically engineered nanoparticle) will be low, a large number of particles present in construction/decorative materials will result in complexity that provides sufficient computing efficiency and energy effectiveness. 

How much intelligence can one squeeze in a single brick? In principle, if one is to trust philosophers, the answer is: every computation one can think off~\cite{RN619}. In the appendix of his cited book~\cite{RN619}, Hillary Putnam conceived a thought experiment to challenge the idea that the state of the mind can be described as a state of a computing automaton. To show that the ability to compute does not automatically constitute intelligence, Putnam provided an elaborate construction how to compute with a rock, and the computing brick is not far away. Naturally, Putnam's arguments have been severely criticised on the account that to turn a rock in a computer one would need to equip it with computationally heavy interface layer. There are other hidden paradoxes associated with that idea, that are not that obvious. For example, how come that we have to struggle so much to achieve computation if such a simple object as a rock or a brick has an intrinsic ability to compute? As it turns out, to resolve such paradoxes, the right question to ask is what is the computation that can be performed {\em naturally} by an object. The turning point where the complexity of the interface layer overpowers the complexity of the system marks the natural boundary of what the system can compute naturally~\cite{RN640}. What is this naturally boundary for a brick? Without a dedicated engineering effort, a brick cannot compute, obviously. However, we argue that by using the materials discussed above, a brick can compute not only in the Putnam's sense (as any rock would do), but as a practical computing device, if used as a reservoir computer.

A reservoir computer consists of two parts, a dynamical system that converts inputs into the states of the reservoir, and the readout layer that is used to analyse the state of the reservoir. The act of computation is the transformation of the external input into the internal state of the reservoir. This step is carried out automatically by the system. In principle, one should be able to use every dynamical system this way. However, by using rigorous mathematical arguments it has been proven that only those systems that separate inputs can operate like that. As a rule of thumb, the more complex the system, the more likely it is a good reservoir. Naturally, there is a plethora of possibilities to achieve such complexity in the design of the computing building. 

For example, the interconnected network of components embedded into the building constitutes a complex dynamical system. This system can be used as a reservoir in order to achieve reservoir computing. If one could construct such buildings, then a great deal of information could be pre-processed by the building before this information is sent to a central unit for analysis and decision making, and the building would essentially function as an intelligent, cognitive information processing substrate. Embedded intelligence could be used for both {\em in situ} and real-time computation. The key benefits of such reservoir computing are indeed numerous: 
\begin{itemize}
    \item 
    The computing is neuromorphic by construction, which implies that the system would be robust and it would function even if part of it is damaged. 
    \item Noise-tolerance: occasional fluctuations would be ignored.
    \item Ability to generalise: any neuromorphic architecture does it, and if a genuinely different conditions are exhibited, the system would respond in a reasonable way.  
    \item Information bandwidth (and energy) reduction: due to the pre-processing, the information bandwidth (and energy consumption) can be significantly reduced.
\end{itemize}

Developing a theoretical mechanistic understanding regarding how a single brick can respond to the external stimuli, and how to interpret these responses for information processing purposes, is not an easy task. What are the challenges? Very likely, theoretical insights are only possibly through virtual experiments, where the computing operation of bricks is simulated digitally. To do this, we firstly need to develop  a dynamical model of the amorphous material that the brick is made off. Once this is done and the principles are clear, ideally we can develop a circuit analogue of the previously developed model.

What should an adequate model look like? The materials will respond to environmental influences, e.g. light, temperature, mechanical stresses. These stimuli will very likely be converted into electrical signals in the material. The key challenges are to predict 
\begin{itemize}
\item how these signals propagate through the brick, and 
\item how to use them for computation (e.g. sensing). 
\end{itemize}
Should it be possible to understand the interplay between the brick design and its information processing features, then one could identify the most optimal designs with practical relevance for architectural purposes. %
%

Developing an adequate amorphous material model might be the biggest theoretical and, indeed, practical challenge. Once the liquid concrete hardens, it forms at random a network of linked objects that conduct, and potentially modify/tune electrical current. This amorphous material can be modelled as a random network of electronic components, but the challenge is to suggest an appropriate topology for the network and the components. One can envision a series of models with an increasing degree of complexity.

As a first approximation, it might be reasonable to  consider memristor-like components only, and assume the nearest neighbour architecture. If one could infer from the experiments  the probability distribution function that governs the shape of the network one could perform realistic simulations. Given the fact that a brick is a macroscopic object,  statistical physics approaches naturally suggest themselves as a modelling technique. By using the models one can extract the typical response behaviour that is stable across all network realizations, and fit to the experimental data. The model could be further augmented with an additional layer of detail, by e.g. considering charge blockade effects. As the readout layer it is natural to use  microprocessor and embed it in the brick. 

A major challenge is to model environmental influences. Sure, one can model these as current/voltage sources or as environment-sensitive electronic components, but it is not entirely clear how to represent the appropriate input to the network. For example, while one can develop a model of the brain, it is hard to know how to simulate an experience of seeing a flower; which voltage signal should one assume at the axon that comes from the retina?

The fully parametric model can be used to characterise the single-brick information processing capacity. One can simulate a fully functional brick for a range of information processing applications. Performance metrics in terms of computing ability and speed, power consumption, data rate, etc. of the proposed circuitry can be defined and studied, and help to identify the most optimal brick designs and ideal use cases. This mechanistic understanding is important, since it can aid the design of  experimental protocols, physical and software interface, and communication protocols to implement computation with a single brick.
 
 Some key strategic tasks might involve the evaluation of few selected use cases for practical demonstration. In the process one needs to  weight in factors as practical relevance versus ease of implementation.

\section{Computing walls}
\label{computingwalls}

\begin{figure}[!tbp]
    \centering
    \subfigure[]{
    \includegraphics[width=0.4\textwidth]{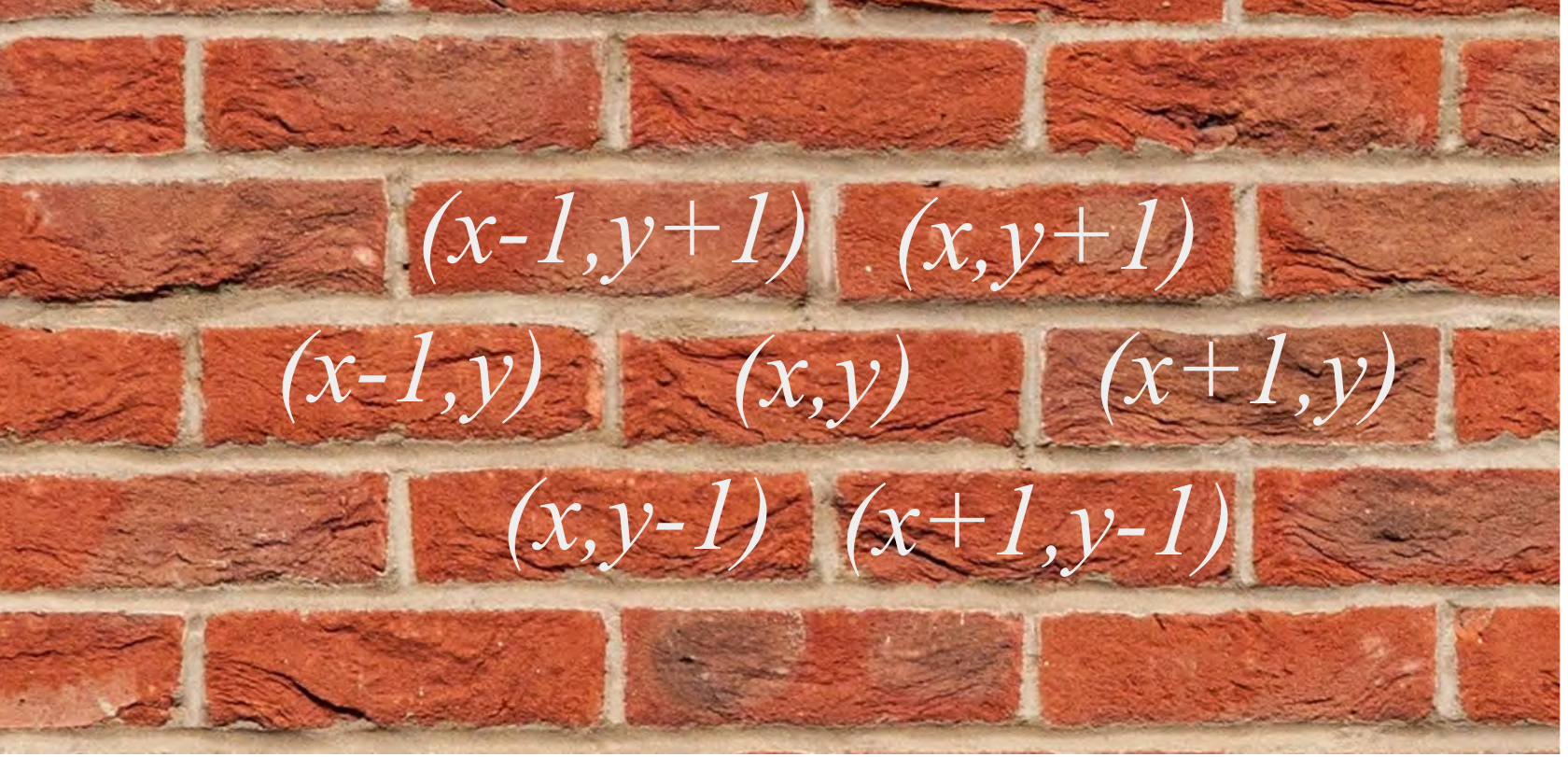}}
    \graphicspath{{figs/23172719/}}
    \subfigure[$t=5$]{\includegraphics[width=0.4\textwidth]{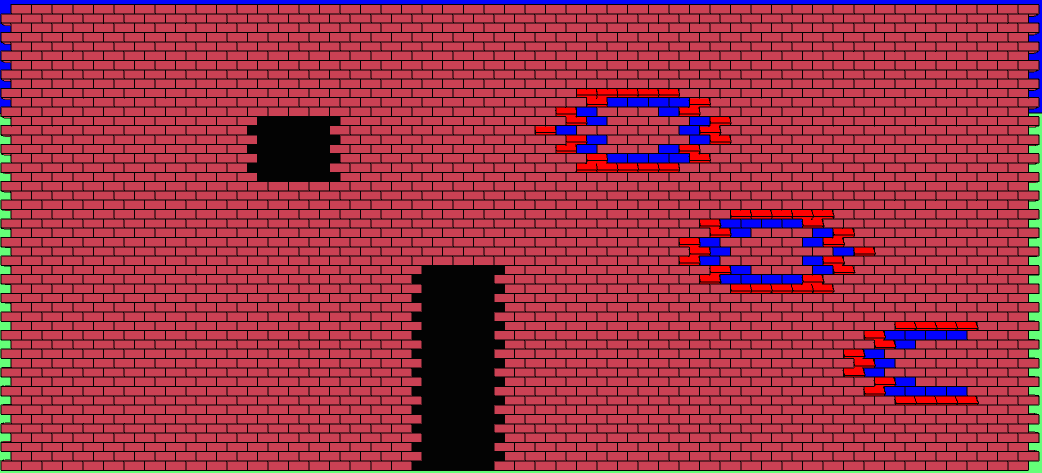}}
    \subfigure[$t=10$]{\includegraphics[width=0.4\textwidth]{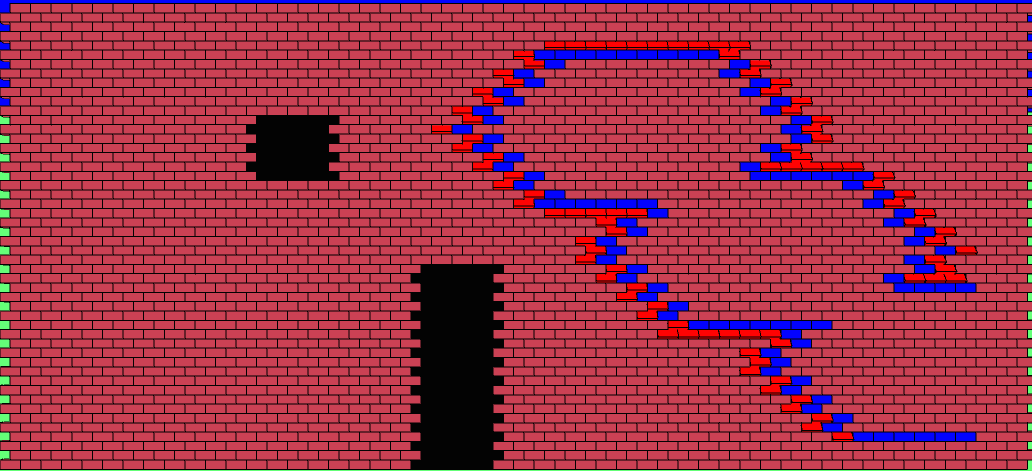}}
    \subfigure[$t=15$]{\includegraphics[width=0.4\textwidth]{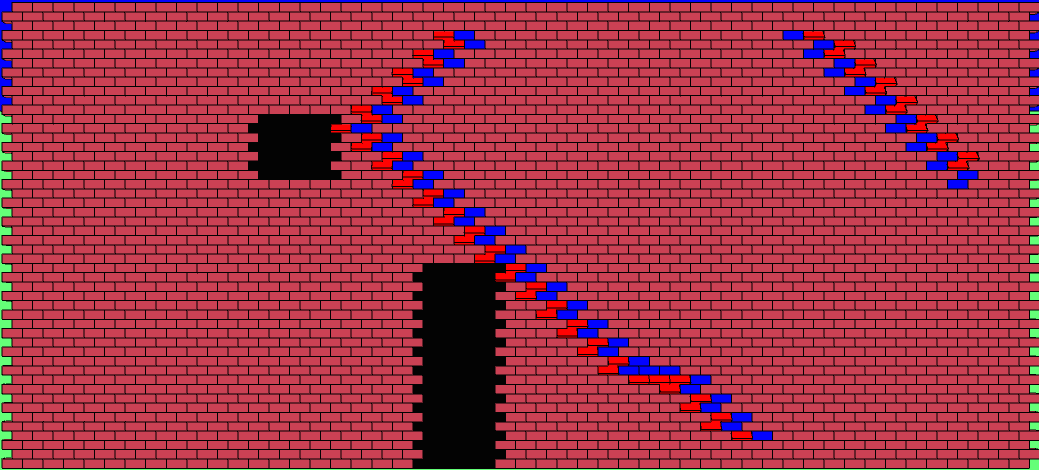}}
    \subfigure[$t=20$]{\includegraphics[width=0.4\textwidth]{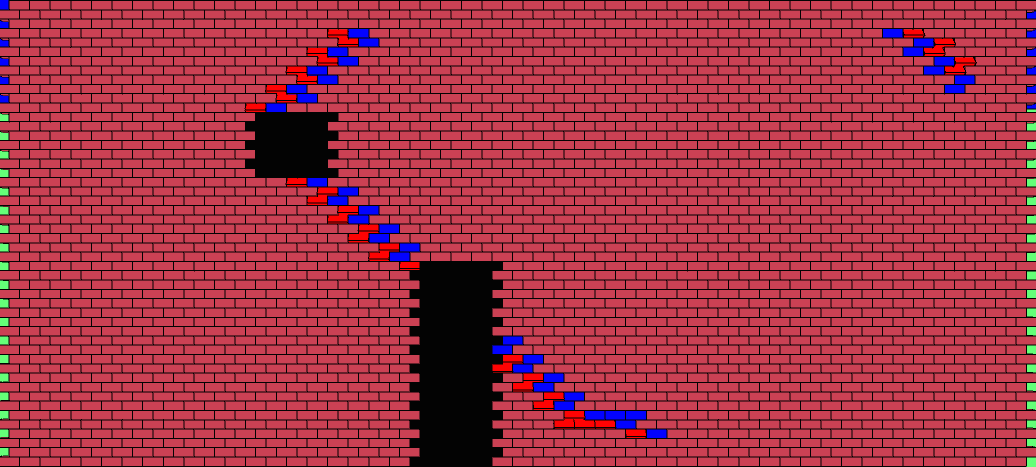}}
    \subfigure[$t=25$]{\includegraphics[width=0.4\textwidth]{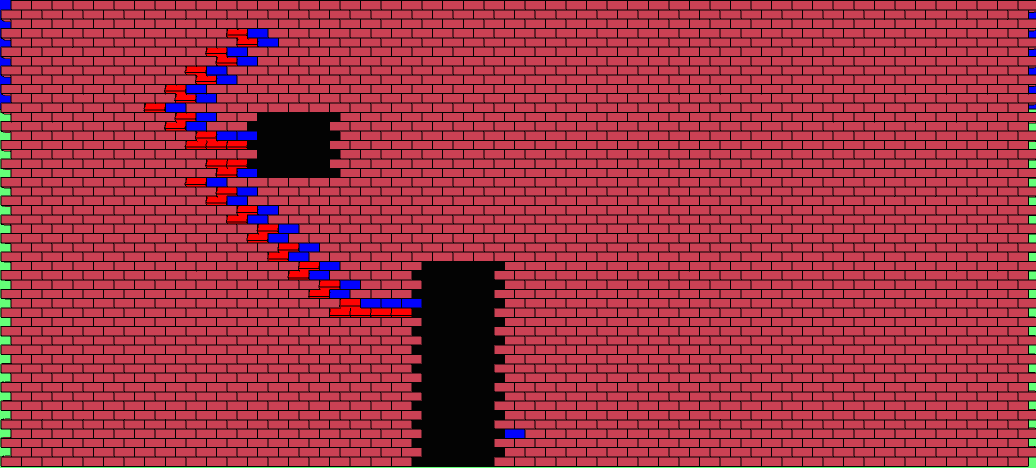}}
    \caption{(a)~Local neighbourhood of a computing brick. Coordinates of neighbours of the brick $(x,y)$ are shown explicitly. (b--f)~Brick wall as an excitable medium. Propagation of the excitation wave-fronts. Excited bricks are red, refractory are blue, resting are brown.}
    \label{fig:neighbourhood}
\end{figure}
\graphicspath{{figs/}}

Drawing from the collection of thoughts and design ideas in the previous section this part of the chapter describes a feasible proposal for construction. We aim to build a wall of computing bricks\footnote{We want to stress again that `brick' is a generalisation of various technological units of a building.} with a small power density that collectively perceives its environment, makes decisions about reconfiguration of the informational contents and performs basic computational primitives with their closest neighbours.

\begin{figure}[!tbp]
\centering
\subfigure[]{\includegraphics[width=0.49\textwidth]{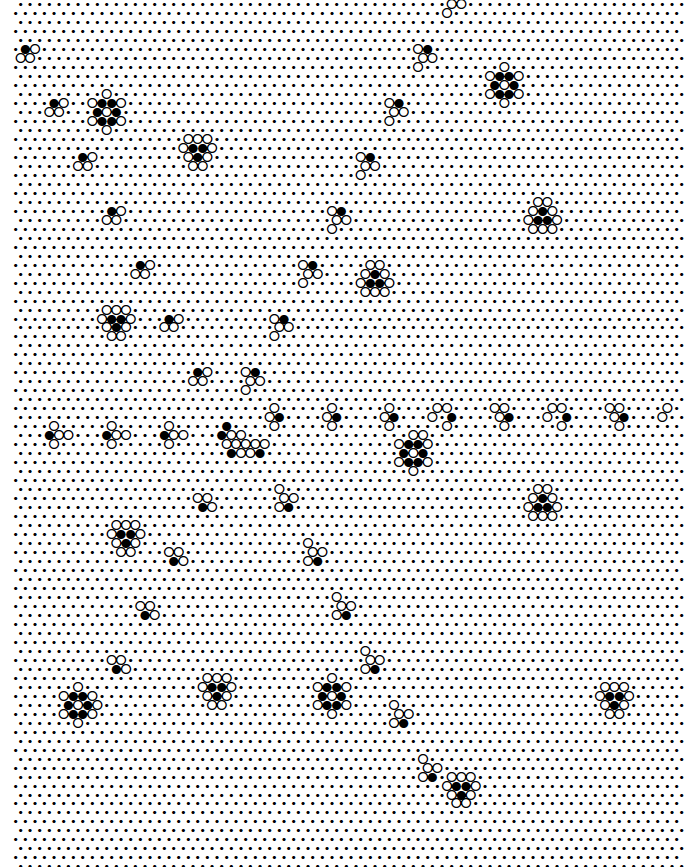}}
\subfigure[]{\includegraphics[width=0.49\textwidth]{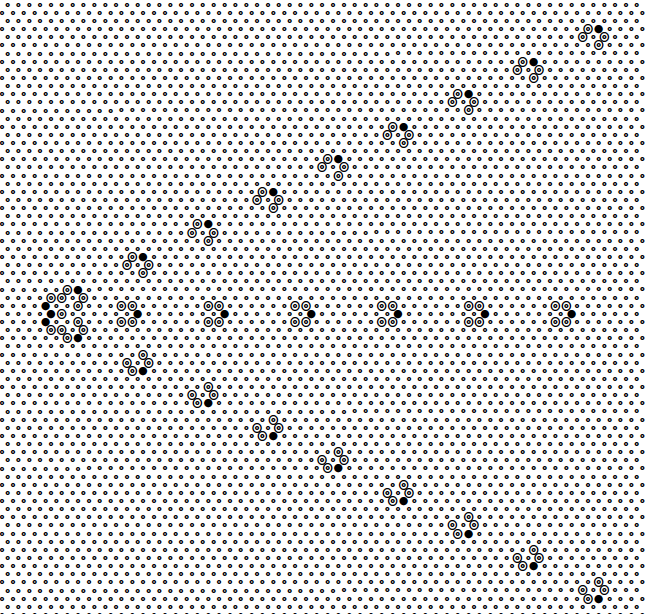}}
\caption{Exemplar configurations of reaction-diffusion automata~\cite{adamatzky2012reaction} with localised transmission of information. (a) A typical quasi-stable configuration of the reaction-diffusion cellular automaton, see rules in~\cite{adamatzky2006computing,adamatzky2012reaction}, which started its
  development in a random initial configuration (with $1/3$ probability of
  each cell-state). Cell-state $I$ (inhibitor) is shown by a black disk,
  cell-state $A$ (activator ) by a circle, and cell-state $S$ (substrate) by a
  dot. We can see there are two types of stationary localisations (glider
  eaters) and a spiral glider-gun, which emits six streams of gliders, with frequency 
  a glider per six time steps in each glider stream.
  (b)~A mobile glider gun travel west and emits three streams of gliders travelling north-east, east and south-east. From~\cite{adamatzky2006computing,adamatzky2012reaction}.
  }
\label{fig:reactiondiffusionautomata}
\end{figure}

A wall can be seen as hexagonal array of processing elements, where each brick, apart of those placed on the edges, has six neighbour (Fig.~\ref{fig:neighbourhood}a). The wall therefore can be abstracted as a hexagonal cellular automaton. This is an array of finite-state machines, or cells. Each cell updates its state in a discrete time, depending on states of its closest neighbours. An example is shown in (Fig.~\ref{fig:neighbourhood}b--f). This is an excitable cellular automaton. Each brick-cell takes three states: resting, excited and refractory. In the automaton illustrated, a resting brick becomes excited if at least one of its six neighbours is excited. An excited brick moves to refractory state and then to resting state unconditionally, i.e. independently on states of its neighbours. In this example, information can be transmitted along a building's wall by the wave of excitation. This is a broadcasting, or one-to-all transmission. By adjusting rules of brick-cell transition, as e.g. demonstrated in~\cite{adamatzky2006computing,adamatzky2012reaction} we can achieve localised transmission of information. In this case, `quanta' of information are represented by compact travelling patterns of activator, or excited, states which propagate in a predetermined direction whilst conversing the velocity vector and shape. See examples in Fig.~\ref{fig:reactiondiffusionautomata}.

The key features of the computing wall are 
\begin{itemize}
\item parallelism and multitasking: each brick senses several states of environment and its neighbours, processes several tasks at a time; 
\item distributed sensing: each brick is a massive distributed sensor of millions of nano- and micro-scale computing elements (smart paint and fictionalised cement)
\item changes in luminosity and temperature affect thousands of macro-computing elements;  
\item task flexibility: the array of computing-bricks  is dynamically reprogrammed to switch between tasks; 
\item fault tolerance: the modular structure of the wall allows operation despite faults, scalability, lower maintenance costs. 
\end{itemize}

The bricks will be made of and coated with smart materials, they will implement sensory fusion of data collected by smart materials, perceive results of low-level information processing undertaken by the smart materials and do high-level computation.

The bricks will be equipped with high-level processing units (CPU, memory, analog and digital input and outputs). The units will be gathering outputs form the low-level computing materials, analysing the data and exchanging the data with processing units in six neighbouring building blocks. The task will deal with evaluating parameters of building blocks, outsourcing or designing in the house elementary processing units, undertaking functionality and structural integrity tests. The brick might be supplied with photovoltaic panels (Fig.~\ref{fig:PVpanel}), to provide an autonomous power supply. In the result we will get manufacturing protocols and prototypes of brick-processors, from which the computing house will be built.  Benchmarking will include processing power, heat generation (and potential associated cooling problems), power consumption, water tightness, frost resistance, mechanical strength.
 
Direct communication between bricks is possible, if e.g. a conductive mortar is uses, however developing a wireless communication network might be more practical. To develop such a network we should analyse a feasible range of wireless transmitters and receivers, evaluate antenna size as a function of frequency, communication neighbourhood size, develop a hardware prototype (just processing units) of the network, analyse communication protocols and address security issues. Particular attention should be paid to fault tolerance and energy saving in the communication network, routing protocols in scenarios of several failed units and distributed storage of information and allocation of concurrent tasks. 

\begin{figure}[!tbp]
    \centering
    \includegraphics[width=\textwidth]{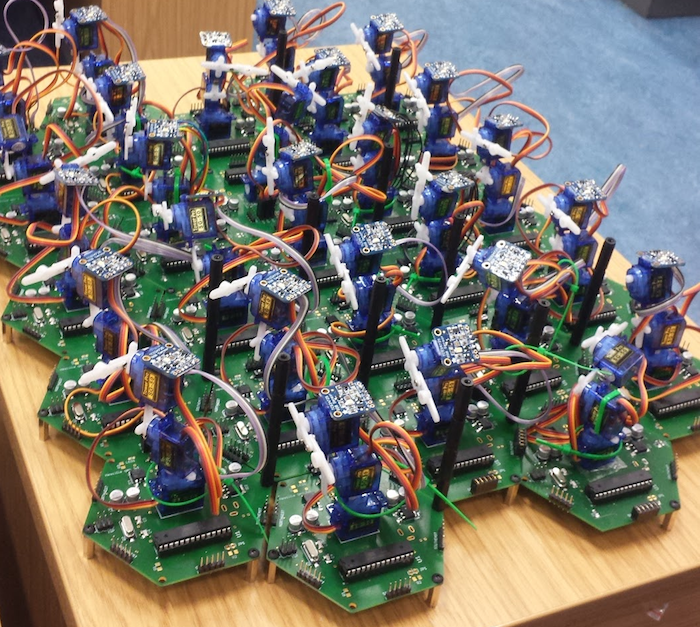}
    \caption{Hexagonal array of processors and manipulators. From~\cite{whiting2018parallel}.}
    \label{fig:hexagonalprototype}
\end{figure}

A physical prototype could be a 3~m by 3~m wall of 600 computing bricks, with high-level processing units embedded, saturated with carbon nanostructures sensitive towards pressure, sounds and coated with photoresponsive smart paints. This massive-parallel array of multiscale computing units will have parallel optical and mechanical inputs and parallel outputs. The computational potential of the array can be benchmarked on the following tasks: computational geometry (plane tessellations or Voronoi diagrams), image processing (contouring, dilation, erosion, detection of features, shape restoration), pattern recognition (via perceptron like models and learning neural networks), optimisation on graphs. There are cellular automata implementations of the solutions of these problems, see e.g. \cite{adamatzky1996voronoi,rosin2006training,popovici2002cellular,rosin2010image,rosin2014cellular,raghavan1993cellular,adamatzky1996computation,behring2001algorithm}, therefore practical implementation would be a matter of technicalities.

As an early proof of concepts we consider our design of a hexagonal array of processors and manipulators capable for object recognition, sorting and transportation (Fig.~\ref{fig:hexagonalprototype}). The processor boards share power and communicate locally~\cite{whiting2018parallel}. Each processor has also a colour sensor to recognise different objects. Each actuator is controlled by a microprocessor. The microprocessor in each cell can detect if an object was present above the cell and the object's colour, control actuation, communicate the cell's state to its six neighbours. In experiments with this physical prototype we developed efficient algorithms for neighbour communication sharing, and solutions for determining shapes objects, based on a configuration of an object's corners.

\begin{figure}[!tbp]
    \centering
    \subfigure[]{
    \includegraphics[width = 0.22\textwidth]{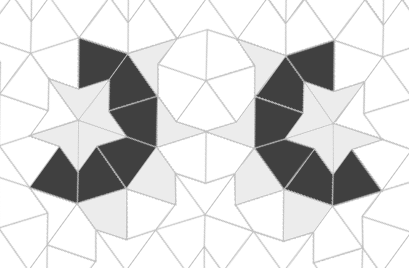}~~ 
\includegraphics[width = 0.22\textwidth]{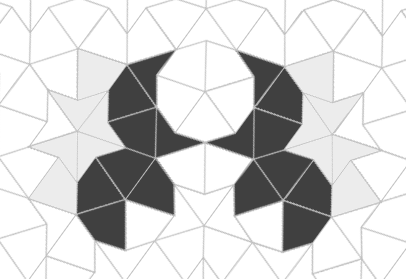}~~
\includegraphics[width = 0.22\textwidth]{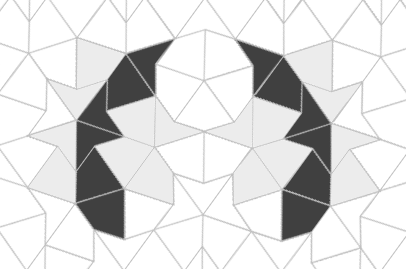}~~
\includegraphics[width = 0.23\textwidth]{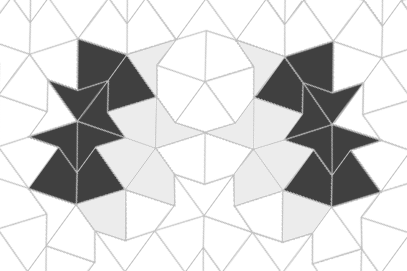}
}
\subfigure[]{\includegraphics[width= 0.7\textwidth]{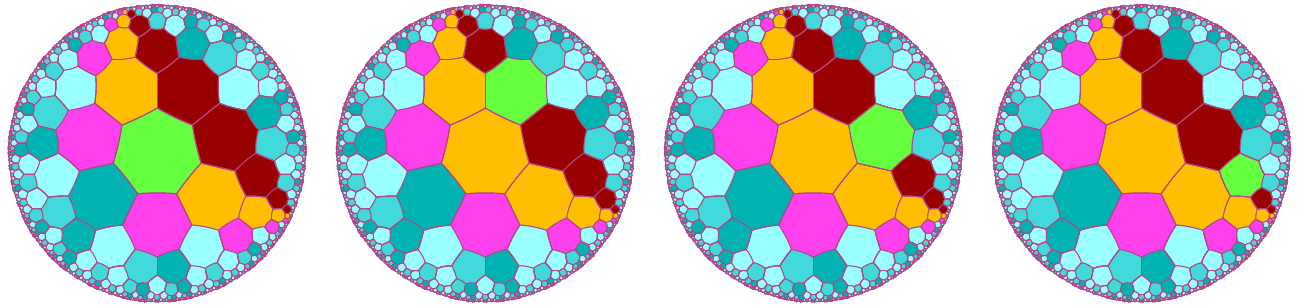}}
    \caption{
    (a)~Configurations of a stationary oscillator in Game of Life like cellular automaton on Penrose tiling discovered by Owens and Stepney~\cite{owens2010investigations,stepney2016art}.
    (b)~Configurations of a strongly universal cellular automaton on a heptagrid, designed by Margenstern~\cite{margenstern2011universal}. The configuration shows detection of the halting state of the simulated Turing machine. From~\cite{margenstern2016hyperbolic}.
    }
    \label{fig:tilings}
\end{figure}

Hexagonal arrangement of bricks is just an example. Indeed, a wide variety of computing architecture, even those supporting cellular automaton algorithms, can be realised on various classes of tilings.
Thus, a Game of Life cellular automaton rules~\cite{adamatzky2010game} can be run on Penrose tilings~\cite{owens2010investigations}. An example of a period four oscillator is shown in Fig.~\ref{fig:tilings}a:  live cells are shown in black, cells that are always dead are white, and dead cells to be alive next step are grey. A series of universal cellular automata on a hyperbolic plane, heptagrid and dodecagrid tiling, are designed in~\cite{margenstern2000new,margenstern2013small,margenstern2011universal}. An example of evolution of a strongly universal cellular automaton is shown in Fig.~\ref{fig:tilings}a, at the moment shown the automaton detects halting of the simulated Turing machine.

\section{Computing architecture in buildings and urban systems}
\label{computingarchitecture}

`Smart materials', `physical computing' and `interacting surfaces' are still rare goods in architectural design and construction, but have together with the rise of material intelligence in natural materials, gained increasing relevance. Such `thinking materials' bring advantages in performance monitoring or function  and as enablers for social interaction, the space and user-architecture-relationship~\cite{nabil2017interactive,ahlquist2017multisensory}. Programmed materials, e.g., through simple operations such as slicing, or in composite with cooperating materials, have been tested in experimental architectural design studios and labs over the last decade. Material sciences and architecture are beginning to merge~\cite{bechthold2017materials,addington2012smart}.

\begin{figure}[!tbp]
    \centering
    \subfigure[]{\includegraphics[width=0.9\textwidth]{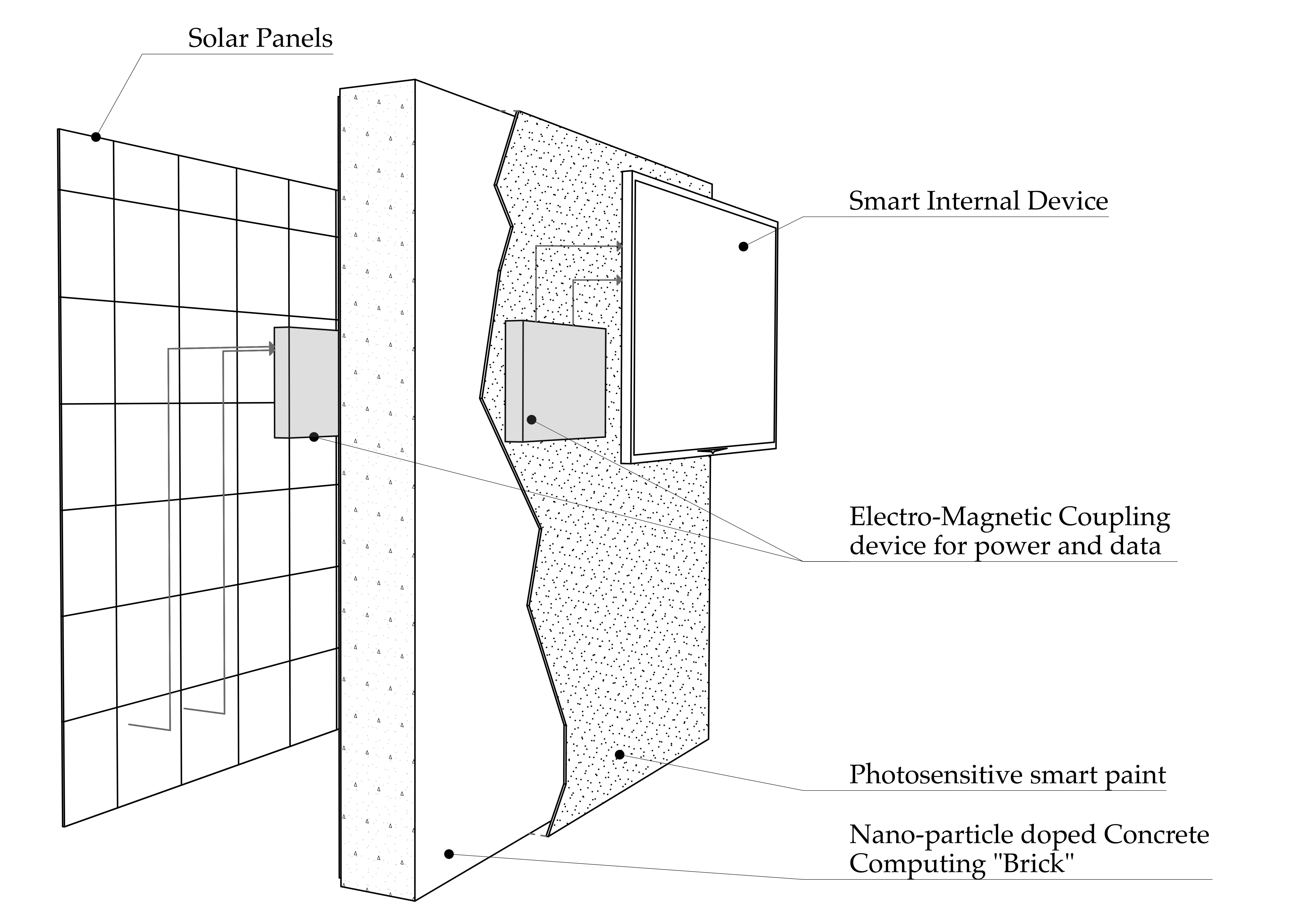}}
    \subfigure[]{\includegraphics[width=0.9\textwidth]{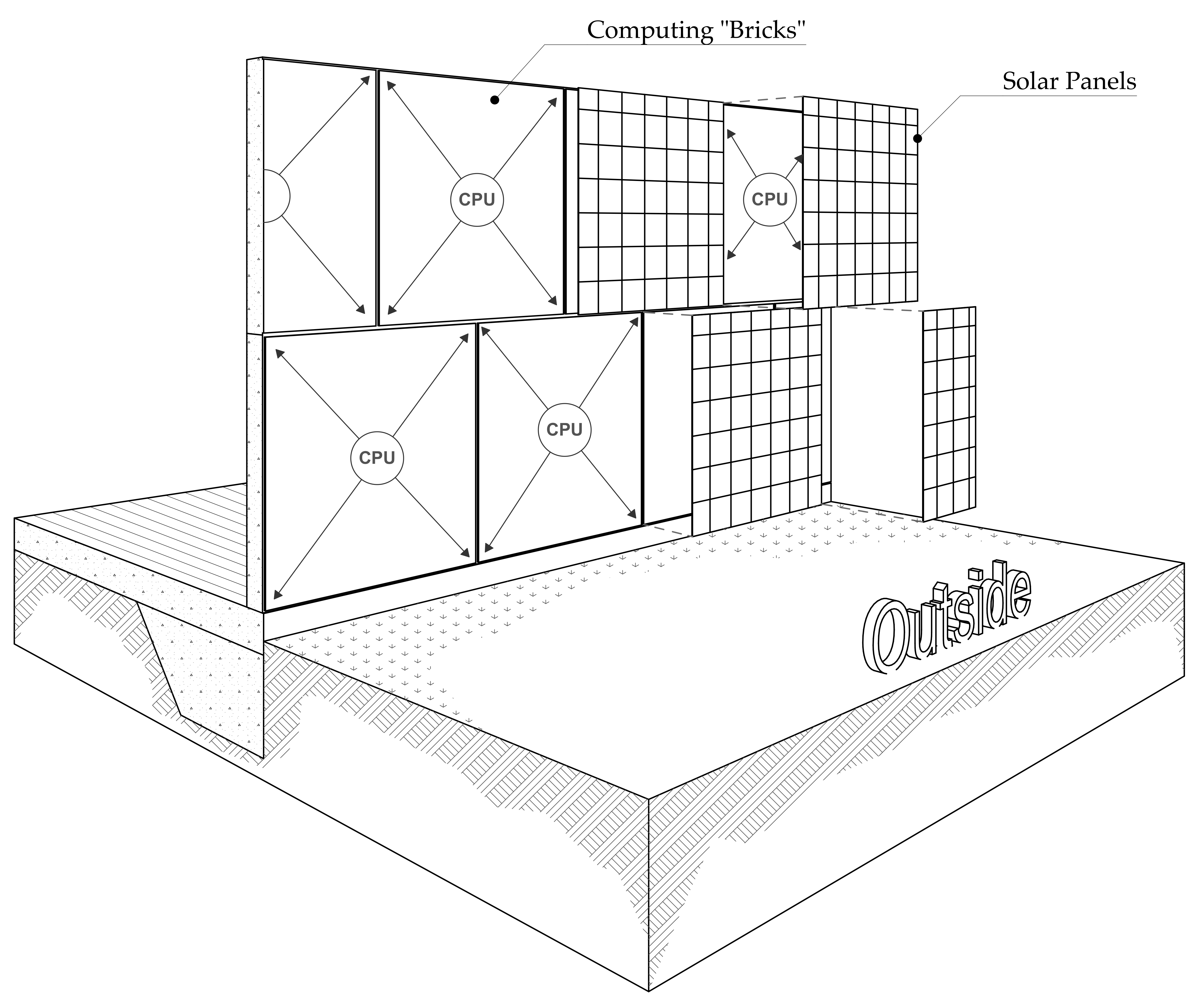}}
    \caption{Power supply to computing building. (a)~Solar panel powered computing brick. (b)~An assembly of bricks and panels.}
    \label{fig:PVpanel}
\end{figure}

A number of architecture schools, scholars and researchers have shifted from designing forms with phenomenological sensing to designing reactive, responsive and interactive surfaces as well as kinetic components and materials featuring a combination of material intelligence and adaptive mechanisms. The implementation of machine learning, deep learning and interconnected sensors so far is in its embryonic age; nevertheless an approach to conceptualise neural networks for architecture is well on its way. Material sciences and physical computing in architectural design are relatively new terrains for architects, educational institutions, building regulations, governments and last but not least the user of a building, the human–and soon also humanoid and robotic agents and instances.
This section focuses on the development of  intelligent buildings by including the computing concrete for bricks in vertical and horizontal pre-fabricated components, walls (Fig.~\ref{fig:PVpanel}).

\begin{figure}[!tbp]
    \centering
    \includegraphics[angle=90,height=0.97\textheight]{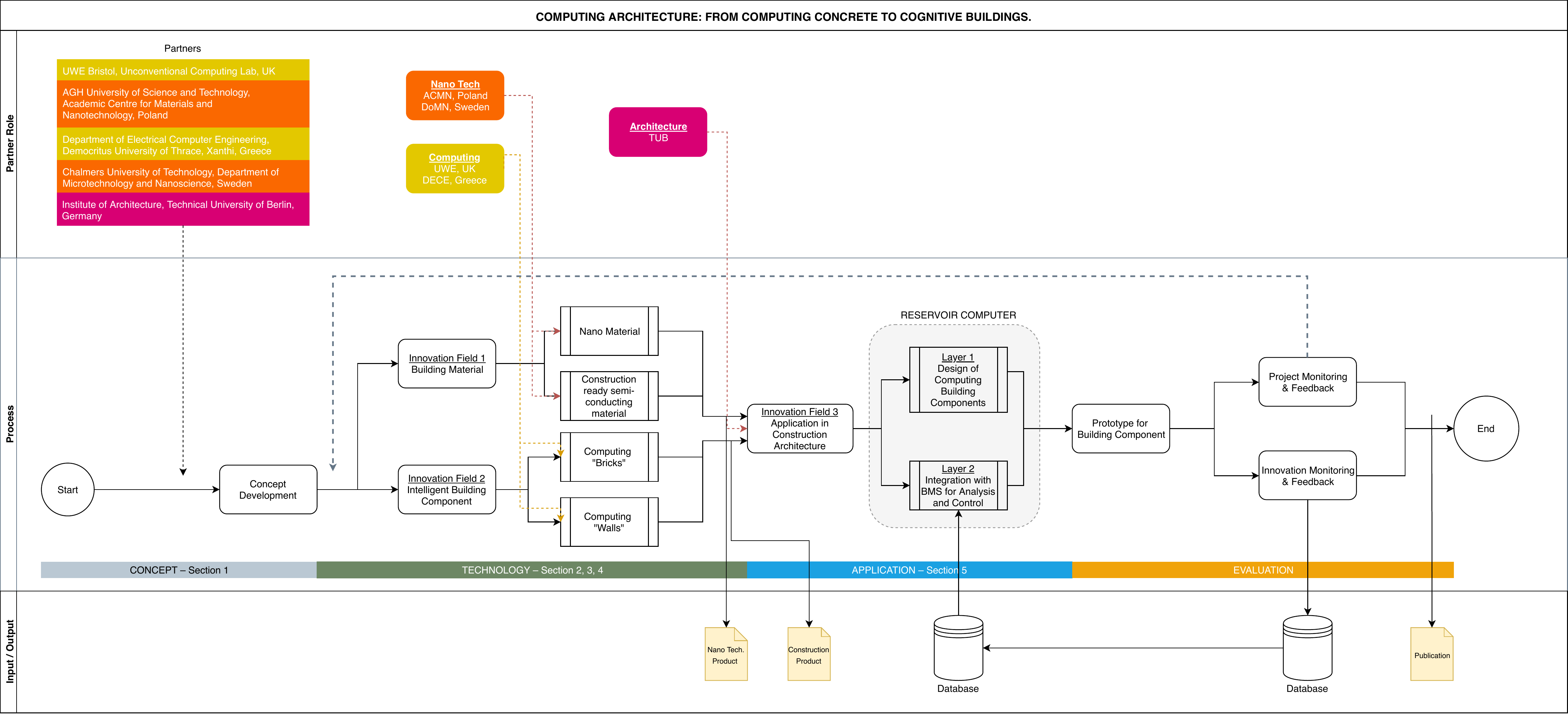}
    \caption{Development and design process of a computing building component.}
    \label{fig:Process}
\end{figure}

The design process of a `Computing Architecture', a brick, a component, a building or a city is highly interdisciplinary and affords a good understanding from each discipline of the other disciplines methods, scale of accuracy, terminology and time required for discipline specific processes. Figure~\ref{fig:Process} shows the four different steps from concept development (what should the material focused computational architecture do) via technological challenges and the development of a smart composite material, the application as building architecture and reservoir computer and finally the implementation in the built environment. Throughout the process multiple feedback loops steer the final product, whereby the demands on the final product, its parameters for success are already being considered in the first step of the concept development. The means of production of the final product --– here a parametrically designed and digitally manufactured building component --- and its function in the domain architecture set a number of parameters in stages 1,2 and 3. In a testing phase art the end of the process we monitor the prototypes performance against a selection of parameters and benchmarks (including human comfort of the occupant of a computing building) in order to feedback into next iteration of the design process. A method based on cybernetics that we developed as `closed-loop digital design process'.

By embedding computing circuits into low-level building materials we will lay foundations for the design of an environment-sensitive computing infrastructure deeply embedded into a body plan for a new building typology. Once applied and integrated in the urban environment, the novel building material offers an unforeseen plethora of applications, increased robustness, sustainability and a hence a new strategy for human-centred design. 
The assembly-instructions for construction would be partly programmed into the material itself, directing digital industrial pre-fabrication methods to arrive at the desired product. The architect becomes partly designer of form and beauty and partly a designer of a construction process. Depending on the advancement of the computational material, the material itself could design the program for self-formation and assembly. Different to known materials that require external force for assembly, advanced computing materials shows the behaviour of what we call `auto-morphological assembly behaviour'. Improved building industry service-and product integration developed during the previous decade known as the age of industry 4.0~\cite{lasi2014industry} laid the foundations for they process. The building typology we are proposing is a cyber-physical system in its material sense and also in the way it is constructed. We envisage the components as information transfer interfaces post-construction and as information carrier pre- and during the construction process of the building.  The computing building materials will allow for faster, improved design and design-to-construction process~\cite{lee2014cyber}.

The application of  photosensitive smart paint on the interior --– and possibly exterior surfaces of buildings at ground floor level --– elevates the physical interface from one room to another or from inside to outside, from a passive to an active part of the building. The proposed materials (doped concrete plus smart paint) are a new approach to architectural construction. Applying smart materials in building architecture in combination with an IoT network has a number of benefits for the building and its occupants.

The architectural prototype we are proposing at this stage has been described as a 3$\times$3~m wall made of nearly 600 standard bricks or a wall constructed of pre-fabricated slabs. On the architectural level we push the physical prototype to a design proposal tested through simulations and physical modelling. The architectural design includes on the one hand material intelligence and on the other the ability to kinetically move slightly in order to advance and compliment the digital computation (enabled through the reservoir computer) with analogue computational (enabled through local material response to the environment. The design solution in combination with sensory digital intelligence for performance evaluation and data collection aims at an adaptable, truly intelligent building, similar to a natural organism. 
One part of the concept is to investigate wireless strategies for buildings and cities and living environments. Another is to investigate new architectural topologies and how they affect human society in an era post-digital towards computational, biological and organisational. We will develop a testing ground for urban communication through principles of the Internet of Things –-- combining physical prototyping and digital simulation/modelling. Detailed objectives before construction include \cite{mohamed2017smart}
\begin{itemize}
\item interface computational materials (concrete, paint and other mineral-based material) with the design processes of a building; 
\item increase sustainable building design; 
\item address the rapid increase in world population necessity for housing and sustainability; 
\item address the relevant industries (building industry and housing associations) and markets for industrial pre-fabricated building components; 
\item developing new efficient and effective tools for building design and performance (qualitative and quantitative), including scripting tools and data. 
\end{itemize}

The objects after construction would be 
\begin{itemize}
\item self-diagnosis (structural failure)
\item energy efficiency
\item adaptability to light and darkness – further development of smart light
\item pattern recognition leading to increased safety 
\item  assistance for bodily or mentally impaired humans 
\end{itemize}

The last objective has been tested in environments for blind people in an outdoor way-finding system~\cite {emerson2017outdoor}. Emerson~\cite {emerson2017outdoor} mentions a project developed by Ohio State University testing smart outdoor paint for visually impaired pedestrians, equipped with a specifically designed cane. In this respect in architecture we could be looking at the communication between building and human resulting in a mutual interactive relationship. Equipped with a smart external device, such as a smart phone, or a smart internal device, such as a microchip implanted under the skin, the human would directly instruct, communicate or teach the building --- and vice versa. A future scenario draws high potential for understanding human behaviour in a specific building and more efficient and improved monitoring of performance, especially failures in parts of the building. 
The first related to machine learning to benefit the occupant, the latter relates to machine learning to benefit the building fabric and function. Every building block will monitor its own states and also will be aware about states of its neighbours. A feature especially relevant in un-manned or hardly manned buildings such as data-centres, power stations or large bridges.


Computational architecture in architecture and design stems from the 1950s, where human-computer interaction was first explored to assist architectural design and to make buildings that could react, interact and somewhat think. Cybernetic architecture as the first strand of computational architecture had been pushed to the extend that cities were envisaged to be super-cities that would operate like super-computers. Concepts known as architecture of utopia developed primarily in the 1960s and 1970s by Superstudio or Archizoom and discussed by Tafuri, Rowe and others circulated around the discussion of technology, society and space. `No-Stop City' by Archizoom was pioneering the idea of an architecture that could grow and continue without boundaries or hierarchies; its repetitive grid reminded at the circuit board of a computer. It suggested that advanced technology would destroy the modern city based on centralization~\cite{contandriopoulos2013architecture,scott2001architecture,stauffer2002utopian}.

Applying a regular grid of programmed micro-computers gives breeding ground for an architecture as network. Similarly to the brain it is not only an idea or utopia, but a material and immaterial phenomenon throughout the history of architecture and urban design.

Architectural design on a building level describes the process in which an architect achieves a solution to combine spatial and functional requirements, with user requirements, external parameters (climate, noise, urban context), material choice (structural, aesthetic, cultural, ecological, economical), fabrication method, construction method, performance measurement and sustainability~\cite{hevner2010design,simon20164}. Design by making, a standard research method in architecture will be applied. We will use computational tools for surface topography and structural optimisation, material reduction and tool instruction. Architectural design or urban planning will utilise generative scripting software for bottom up form-finding. Digitally driven tools will allow us to negotiation between traditional top-down planning and bottom-up self-organisation. Open source software will be used to generate form and surface topography of the physical architectural building prototype of a wall. The building prototype is developed to serve `Direct Digital Design to Assembly' for an efficient work-flow. The computing walls will be designed to fit CAM digitally prefabrication processes.  

The implementation will be dealt with at several scales. At an architectural scale we will design a prototype using scripting (algorithms based on neural behaviour), building information modelling packages including material intelligence, the integration of climatic data and physical material property data, genetic algorithms. At the building scale we will construct a digitally manufactured (robotic manufacturing and stereo-lithography printing), intelligent component prototype for a building; surface topography will change in accordance to the `behaviour' of the building. At this state a particular attention should be paid to investigate  spatial impact and quality through virtual and augmented reality. At the urban scale we will develop and test  design strategies for industrial pre-fabrication integrating. The urban design strategies will be developed through simulations and virtual reality for future sustainable and self-organising cities, driven by material intelligence and neuro-network distributed computing.

\section{Discussion}
\label{discussion}

We presented our vision of computing architectures. Functional nanoparticles and fibres are mixed in a concrete. Blocks made of this concrete are capable for reservoir computing with potentially thousands of inputs and outputs. Each block is supplied with a microprocessor for input and output interface and communication with neighbouring blocks. A wall made of the blocks is a massively parallel array processor, while each block is a massively parallel reservoir computer. Practical implementation of the idea would involve a high-disciplinary task force. In computer science, we will need to employ a theory of embedded computation with amorphous substrates and to identifying possible advanced sensing applications with amorphous reservoirs and the sensing capacity information of deeply embedded sensing systems in architecture. From nanotechnology and novel construction materials we will need new nano-materials for \emph{in-materio} computing and data acquisition (sensing), development of composite materials combining mechanical properties of concrete and functional properties of nano-materials, new hybrid, stimuli-responsive materials sensitive to light, electric currents, vibrations, sound and humidity. The field of  architecture and building environment will give us new building operations monitoring technology through integrated intelligence in the building material, integration of existing computational design strategies into an intelligent building material.

\section{Acknowledgements}
\label{Acknowledgements}
KS and DP acknowledge the financial support from the National Science Centre (Poland) within the OPUS project, contract No. UMO-2015/17/B/ST8/01783 and from Polish Ministry of Science and Higher Education. 
Authors thank Neil Phillips for precious technical discussions and  Julian F. Miller for helping to improve the paper further.

\bibliographystyle{plain}
\bibliography{biblio}

\begin{thebibliography}{10}

\bibitem{adamatzky2013memristor}
A.~Adamatzky and L.~Chua.
\newblock {\em Memristor Networks}.
\newblock SpringerLink : B{\"u}cher. Springer International Publishing, 2013.

\bibitem{adamatzky1996computation}
AI~Adamatzky.
\newblock Computation of shortest path in cellular automata.
\newblock {\em Mathematical and Computer Modelling}, 23(4):105--113, 1996.

\bibitem{adamatzky1996voronoi}
AI~Adamatzky.
\newblock Voronoi-like partition of lattice in cellular automata.
\newblock {\em Mathematical and Computer Modelling}, 23(4):51--66, 1996.

\bibitem{adamatzky2010game}
Andrew Adamatzky, editor.
\newblock {\em Game of life cellular automata}.
\newblock Springer, 2010.

\bibitem{adamatzky2012reaction}
Andrew Adamatzky.
\newblock {\em Reaction-Diffusion Automata: Phenomenology, Localisations,
  Computation}, volume~1.
\newblock Springer Science \& Business Media, 2012.

\bibitem{adamatzky2006computing}
Andrew Adamatzky and Andrew Wuensche.
\newblock Computing in spiral rule reaction-diffusion hexagonal cellular
  automaton.
\newblock {\em Complex Systems}, 16(4):277--298, 2006.

\bibitem{addington2012smart}
Michelle Addington and Daniel Schodek.
\newblock {\em Smart Materials and Technologies in Architecture: For the
  Architecture and Design Professions}.
\newblock Routledge, 2012.

\bibitem{ahlquist2017multisensory}
Sean Ahlquist, Leah Ketcheson, and Constanza Colombi.
\newblock Multisensory architecture: The dynamic interplay of environment,
  movement and social function.
\newblock {\em Architectural Design}, 87(2):90--99, 2017.

\bibitem{alam2012review}
Muhammad~Raisul Alam, Mamun Bin~Ibne Reaz, and Mohd Alauddin~Mohd Ali.
\newblock A review of smart homes—past, present, and future.
\newblock {\em IEEE Transactions on Systems, Man, and Cybernetics, Part C
  (Applications and Reviews)}, 42(6):1190--1203, 2012.

\bibitem{augusto2006designing}
Juan~Carlos Augusto and Chris~D Nugent.
\newblock {\em Designing smart homes: the role of artificial intelligence},
  volume 4008.
\newblock Springer, 2006.

\bibitem{bechthold2017materials}
Martin Bechthold and James~C Weaver.
\newblock Materials science and architecture.
\newblock {\em Nature Reviews Materials}, 2(12):17082, 2017.

\bibitem{behring2001algorithm}
C~Behring, M~Bracho, M~Castro, and JA~Moreno.
\newblock An algorithm for robot path planning with cellular automata.
\newblock In {\em Theory and practical issues on cellular automata}, pages
  11--19. Springer, 2001.

\bibitem{KS_fuzzy}
A.~Blachecki, J.~Mech-Piskorz, M.~Gajewska, K.~Mech, K.~Pilarczyk, and
  K.~Szaciłowski.
\newblock Organotitania-based nanostructures as a suitable platform for the
  implementation of binary, ternary, and fuzzy logic systems.
\newblock {\em ChemPhysChem}, 18:1798--1810, 2015.

\bibitem{Williams}
Julien Borghetti, Gregory~S. Snider, Philip Kuekes, J.~Joshua Yang, Duncan~R.
  Stewart, and R.~Stanley Williams.
\newblock 'memristive' switches enable 'stateful' logic operations via material
  implication.
\newblock {\em Nature}, 464(7290):873--876, 2010.

\bibitem{broersma2017computational}
Hajo Broersma, Julian~F Miller, and Stefano Nichele.
\newblock Computational matter: Evolving computational functions in nanoscale
  materials.
\newblock In {\em Advances in Unconventional Computing}, pages 397--428.
  Springer, 2017.

\bibitem{brush2018smart}
AJ~Brush, Mike Hazas, and Jeannie Albrecht.
\newblock Smart homes: Undeniable reality or always just around the corner?
\newblock {\em IEEE Pervasive Computing}, (1):82--86, 2018.

\bibitem{burkow2016exploring}
Aleksander~Vognild Burkow.
\newblock Exploring physical reservoir computing using random boolean networks.
\newblock Master's thesis, NTNU, 2016.

\bibitem{chan2008review}
Marie Chan, Daniel Est{\`e}ve, Christophe Escriba, and Eric Campo.
\newblock A review of smart homes—present state and future challenges.
\newblock {\em Computer methods and programs in biomedicine}, 91(1):55--81,
  2008.

\bibitem{chang2011synaptic}
Ting Chang, Sung-Hyun Jo, Kuk-Hwan Kim, Patrick Sheridan, Siddharth Gaba, and
  Wei Lu.
\newblock Synaptic behaviors and modeling of a metal oxide memristive device.
\newblock {\em Applied physics A}, 102(4):857--863, 2011.

\bibitem{contandriopoulos2013architecture}
Christina Contandriopoulos.
\newblock Architecture and utopia in the 21st-century.
\newblock {\em Journal of Architectural Education}, 67(1):3--6, 2013.

\bibitem{dale2017reservoir}
Matthew Dale, Julian~F Miller, and Susan Stepney.
\newblock Reservoir computing as a model for in-materio computing.
\newblock In {\em Advances in Unconventional Computing}, pages 533--571.
  Springer, 2017.

\bibitem{dale2016evolving}
Matthew Dale, Julian~F Miller, Susan Stepney, and Martin~A Trefzer.
\newblock Evolving carbon nanotube reservoir computers.
\newblock In {\em International Conference on Unconventional Computation and
  Natural Computation}, pages 49--61. Springer, 2016.

\bibitem{darianian2008smart}
Mohsen Darianian and Martin~Peter Michael.
\newblock Smart home mobile rfid-based internet-of-things systems and services.
\newblock In {\em Advanced Computer Theory and Engineering, 2008. ICACTE'08.
  International Conference on}, pages 116--120. IEEE, 2008.

\bibitem{emerson2017outdoor}
Robert~Wall Emerson.
\newblock Outdoor wayfinding and navigation for people who are blind: Accessing
  the built environment.
\newblock In {\em International Conference on Universal Access in
  Human-Computer Interaction}, pages 320--334. Springer, 2017.

\bibitem{nanoscale}
S.~Gawęda, A.~Podborska, W.~Macyk, and K.~Szaciłowski.
\newblock Nanoscale optoelectronic switches and logic devices.
\newblock {\em Nanoscale}, 1:299--316, 2009.

\bibitem{hevner2010design}
Alan Hevner and Samir Chatterjee.
\newblock {\em Design research in information systems: theory and practice},
  volume~22.
\newblock Springer Science \& Business Media, 2010.

\bibitem{physical_system}
C.~Horsman, S.~Stepney, R.C. Wagner, and V.~Kendon.
\newblock When does a physical system compute?
\newblock {\em Proc. Royal. Soc. A}, 470:20140182, 2014.

\bibitem{kasabov2016evolving}
Nikola Kasabov, Nathan~Matthew Scott, Enmei Tu, Stefan Marks, Neelava Sengupta,
  Elisa Capecci, Muhaini Othman, Maryam~Gholami Doborjeh, Norhanifah Murli,
  Reggio Hartono, et~al.
\newblock Evolving spatio-temporal data machines based on the neucube
  neuromorphic framework: design methodology and selected applications.
\newblock {\em Neural Networks}, 78:1--14, 2016.

\bibitem{RN697}
Zoran Konkoli.
\newblock {\em On reservoir computing: from mathematical foundations to
  unconventional applications}, volume 1. Theory.
\newblock Springer, 2016.

\bibitem{kuzum2013synaptic}
Duygu Kuzum, Shimeng Yu, and HS~Philip Wong.
\newblock Synaptic electronics: materials, devices and applications.
\newblock {\em Nanotechnology}, 24(38):382001, 2013.

\bibitem{larger2017high}
Laurent Larger, Antonio Bayl{\'o}n-Fuentes, Romain Martinenghi, Vladimir~S
  Udaltsov, Yanne~K Chembo, and Maxime Jacquot.
\newblock High-speed photonic reservoir computing using a time-delay-based
  architecture: Million words per second classification.
\newblock {\em Physical Review X}, 7(1):011015, 2017.

\bibitem{larger2012photonic}
Laurent Larger, Miguel~C Soriano, Daniel Brunner, Lennert Appeltant, Jose~M
  Guti{\'e}rrez, Luis Pesquera, Claudio~R Mirasso, and Ingo Fischer.
\newblock Photonic information processing beyond turing: an optoelectronic
  implementation of reservoir computing.
\newblock {\em Optics express}, 20(3):3241--3249, 2012.

\bibitem{lasi2014industry}
Heiner Lasi, Peter Fettke, Hans-Georg Kemper, Thomas Feld, and Michael
  Hoffmann.
\newblock Industry 4.0.
\newblock {\em Business \& Information Systems Engineering}, 6(4):239--242,
  2014.

\bibitem{lee2014cyber}
Jay Lee, Behrad Bagheri, and Hung-An Kao.
\newblock A cyber-physical systems architecture for industry 4.0-based
  manufacturing systems.
\newblock {\em Manufacturing Letters}, 3:18--23, 2015.

\bibitem{KS_demult}
K.~Lewandowska, A.~Podborska, P.~Kwolek, T.-D. Kim, Lee. K.-S., and
  K.~Szaciłowski.
\newblock Optical signal demultiplexing and conversion in the
  fullerene–oligothiophene–cds system.
\newblock {\em Appl. Surf. Sci.}, 319:285--290, 2014.

\bibitem{li2011smart}
Xu~Li, Rongxing Lu, Xiaohui Liang, Xuemin Shen, Jiming Chen, and Xiaodong Lin.
\newblock Smart community: an internet of things application.
\newblock {\em IEEE Communications Magazine}, 49(11), 2011.

\bibitem{lukovsevivcius2009reservoir}
Mantas Luko{\v{s}}evi{\v{c}}ius and Herbert Jaeger.
\newblock Reservoir computing approaches to recurrent neural network training.
\newblock {\em Computer Science Review}, 3(3):127--149, 2009.

\bibitem{lukovsevivcius2012reservoir}
Mantas Luko{\v{s}}evi{\v{c}}ius, Herbert Jaeger, and Benjamin Schrauwen.
\newblock Reservoir computing trends.
\newblock {\em KI-K{\"u}nstliche Intelligenz}, 26(4):365--371, 2012.

\bibitem{maass2002real}
Wolfgang Maass, Thomas Natschl{\"a}ger, and Henry Markram.
\newblock Real-time computing without stable states: A new framework for neural
  computation based on perturbations.
\newblock {\em Neural computation}, 14(11):2531--2560, 2002.

\bibitem{mandic2001recurrent}
Danilo~P Mandic and Jonathon Chambers.
\newblock {\em Recurrent neural networks for prediction: learning algorithms,
  architectures and stability}.
\newblock John Wiley \& Sons, Inc., 2001.

\bibitem{margenstern2000new}
Maurice Margenstern.
\newblock New tools for cellular automata in the hyperbolic plane.
\newblock {\em Journal of Universal Computer Science}, 6(12):1226--1252, 2000.

\bibitem{margenstern2011universal}
Maurice Margenstern.
\newblock A universal cellular automaton on the heptagrid of the hyperbolic
  plane with four states.
\newblock {\em Theoretical Computer Science}, 412(1-2):33--56, 2011.

\bibitem{margenstern2013small}
Maurice Margenstern.
\newblock {\em Small universal cellular automata in hyperbolic spaces: A
  collection of jewels}, volume~4.
\newblock Springer Science \& Business Media, 2013.

\bibitem{margenstern2016hyperbolic}
Maurice Margenstern.
\newblock Hyperbolic gallery.
\newblock In {\em Designing Beauty: The Art of Cellular Automata}, pages
  65--71. Springer, 2016.

\bibitem{mohamed2017smart}
Abeer Samy~Yousef Mohamed.
\newblock Smart materials innovative technologies in architecture; towards
  innovative design paradigm.
\newblock {\em Energy Procedia}, 115:139--154, 2017.

\bibitem{nabil2017interactive}
Sara Nabil, Thomas Pl{\"o}tz, and David~S Kirk.
\newblock Interactive architecture: Exploring and unwrapping the potentials of
  organic user interfaces.
\newblock In {\em Proceedings of the Eleventh International Conference on
  Tangible, Embedded, and Embodied Interaction}, pages 89--100. ACM, 2017.

\bibitem{owens2010investigations}
Nick Owens and Susan Stepney.
\newblock Investigations of game of life cellular automata rules on penrose
  tilings: Lifetime, ash, and oscillator statistics.
\newblock {\em J. Cellular Automata}, 5(3):207--225, 2010.

\bibitem{Papandroulidakis2017}
G.~Papandroulidakis, I.~Vourkas, A.~Abusleme, G.~C. Sirakoulis, and A.~Rubio.
\newblock Crossbar-based memristive logic-in-memory architecture.
\newblock {\em IEEE Transactions on Nanotechnology}, 16(3):491--501, May 2017.

\bibitem{Papandroulidakis2014}
G.~Papandroulidakis, I.~Vourkas, N.~Vasileiadis, and G.~C. Sirakoulis.
\newblock Boolean logic operations and computing circuits based on memristors.
\newblock {\em IEEE Transactions on Circuits and Systems II: Express Briefs},
  61(12):972--976, Dec 2014.

\bibitem{park2015electronic}
Sangsu Park, Myonglae Chu, Jongin Kim, Jinwoo Noh, Moongu Jeon, Byoung~Hun Lee,
  Hyunsang Hwang, Boreom Lee, and Byung-geun Lee.
\newblock Electronic system with memristive synapses for pattern recognition.
\newblock {\em Scientific reports}, 5:10123, 2015.

\bibitem{KS_neuro}
K.~Pilarczyk, A.~Podborska, M.~Lis, M.~Kawa, Migdał D., and K.~Szaciłowski.
\newblock Synaptic behavior in an optoelectronic device based on
  semiconductor-nanotube hybrid.
\newblock {\em Adv.Electr. Mat.}, 2:1500471, 2016.

\bibitem{KS_sensors}
K.~Pilarczyk, E.~Wlaźlak, D.~Przyczyna, A.~Blachecki, A.~Podborska,
  V.~Anathasiou, Z.~Konkoli, and K.~Szaciłowski.
\newblock Molecules, semiconductors, light and information: Towards future
  sensing and computing paradigms.
\newblock {\em Coord. Chem. Rev.}, 265:22--40, 2017.

\bibitem{KS_sum}
A.~Podborska and K.~Szaciłowski.
\newblock ‘computer-on-a-particle’ devices: Optoelectronic 1:2
  demultiplexer based on nanostructured cadmium sulfide.
\newblock {\em Aust. J. Chem.}, 63:165--168, 2010.

\bibitem{popovici2002cellular}
Adriana Popovici and Dan Popovici.
\newblock Cellular automata in image processing.
\newblock In {\em Fifteenth International Symposium on Mathematical Theory of
  Networks and Systems}, volume~1, pages 1--6. Citeseer, 2002.

\bibitem{RN619}
Hilary Putnam.
\newblock {\em Representation and Reality}.
\newblock MIT Press, Cambridge, 1988.

\bibitem{raghavan1993cellular}
Raghu Raghavan.
\newblock Cellular automata in pattern recognition.
\newblock {\em Information Sciences}, 70(1-2):145--177, 1993.

\bibitem{rosin2014cellular}
Paul Rosin, Andrew Adamatzky, and Xianfang Sun.
\newblock {\em Cellular automata in image processing and geometry}.
\newblock Springer, 2014.

\bibitem{rosin2006training}
Paul~L Rosin.
\newblock Training cellular automata for image processing.
\newblock {\em IEEE transactions on image processing}, 15(7):2076--2087, 2006.

\bibitem{rosin2010image}
Paul~L Rosin.
\newblock Image processing using 3-state cellular automata.
\newblock {\em Computer vision and image understanding}, 114(7):790--802, 2010.

\bibitem{scott2001architecture}
Felicity~D Scott.
\newblock Architecture or techno-utopia.
\newblock {\em Grey Room}, pages 112--126, 2001.

\bibitem{sillin2013theoretical}
Henry~O Sillin, Renato Aguilera, Hsien-Hang Shieh, Audrius~V Avizienis,
  Masakazu Aono, Adam~Z Stieg, and James~K Gimzewski.
\newblock A theoretical and experimental study of neuromorphic atomic switch
  networks for reservoir computing.
\newblock {\em Nanotechnology}, 24(38):384004, 2013.

\bibitem{simon20164}
Herbert~A Simon.
\newblock 4 networks, complexity, models and measures.
\newblock {\em Access, Property and American Urban Space}, page~58, 2016.

\bibitem{stauffer2002utopian}
Marie~Theres Stauffer.
\newblock Utopian reflections, reflected utopias urban designs by archizoom and
  superstudio.
\newblock {\em AA files}, (47):23--36, 2002.

\bibitem{stepney2016art}
Susan Stepney.
\newblock The art of {P}enrose life.
\newblock In Andrew Adamatzky and Genaro Martineze, editors, {\em Designing
  Beauty: The Art of Cellular Automata}, pages 103--109. Springer, 2016.

\bibitem{stepney2012gardening}
Susan Stepney, Ada Diaconescu, Ren{\'e} Doursat, Jean-Louis Giavitto, Taras
  Kowaliw, Ottoline Leyser, Bruce Maclennan, Olivier Michel, Julian Miller,
  Igor Nikolic, et~al.
\newblock Gardening cyber-physical systems.
\newblock In {\em Unconventionnal Computation and Natural Computation
  (UCNC'2012)}, pages 1--1, 2012.

\bibitem{KS_JACS}
K.~Szaciłowski, W.~Macyk, and G.~Stochel.
\newblock Light-driven or and xor programmable chemical logic gates.
\newblock {\em J. Am. Chem.Soc.}, 128:4550--4551, 2006.

\bibitem{ulbricht2011carrier}
Ronald Ulbricht, Euan Hendry, Jie Shan, Tony~F Heinz, and Mischa Bonn.
\newblock Carrier dynamics in semiconductors studied with time-resolved
  terahertz spectroscopy.
\newblock {\em Reviews of Modern Physics}, 83(2):543, 2011.

\bibitem{van2017advances}
Guy Van~der Sande, Daniel Brunner, and Miguel~C Soriano.
\newblock Advances in photonic reservoir computing.
\newblock {\em Nanophotonics}, 6(3):561--576, 2017.

\bibitem{vandoorne2008toward}
Kristof Vandoorne, Wouter Dierckx, Benjamin Schrauwen, David Verstraeten, Roel
  Baets, Peter Bienstman, and Jan Van~Campenhout.
\newblock Toward optical signal processing using photonic reservoir computing.
\newblock {\em Optics express}, 16(15):11182--11192, 2008.

\bibitem{vandoorne2014experimental}
Kristof Vandoorne, Pauline Mechet, Thomas Van~Vaerenbergh, Martin Fiers, Geert
  Morthier, David Verstraeten, Benjamin Schrauwen, Joni Dambre, and Peter
  Bienstman.
\newblock Experimental demonstration of reservoir computing on a silicon
  photonics chip.
\newblock {\em Nature communications}, 5:3541, 2014.

\bibitem{verstraeten2007experimental}
David Verstraeten, Benjamin Schrauwen, Michiel d’Haene, and Dirk Stroobandt.
\newblock An experimental unification of reservoir computing methods.
\newblock {\em Neural networks}, 20(3):391--403, 2007.

\bibitem{vissol2016data}
El{\'e}onore Vissol-Gaudin, Apostolos Kotsialos, Mark~K Massey, Dagou~A Zeze,
  Christopher Pearson, Chris Groves, and Michael~C Petty.
\newblock Data classification using carbon-nanotubes and evolutionary
  algorithms.
\newblock In {\em International Conference on Parallel Problem Solving from
  Nature}, pages 644--654. Springer, 2016.

\bibitem{Memristorbook}
I.~Vourkas and G.Ch. Sirakoulis.
\newblock {\em Memristor-Based Nanoelectronic Computing Circuits and
  Architectures: Foreword by Leon Chua}.
\newblock Emergence, Complexity and Computation. Springer International
  Publishing, 2015.

\bibitem{Ariadne}
I.~Vourkas, D.~Stathis, and G.~C. Sirakoulis.
\newblock Massively parallel analog computing: Ariadne’s thread was made of
  memristors.
\newblock {\em IEEE Transactions on Emerging Topics in Computing},
  6(1):145--155, Jan 2018.

\bibitem{Vourkas2014}
Ioannis Vourkas and Georgios~Ch. Sirakoulis.
\newblock Memristor-based combinational circuits: A design methodology for
  encoders/decoders.
\newblock {\em Microelectronics Journal}, 45(1):59 -- 70, 2014.

\bibitem{Vourkas2014plus}
Ioannis Vourkas and Georgios~Ch. Sirakoulis.
\newblock On the generalization of composite memristive network structures for
  computational analog/digital circuits and systems.
\newblock {\em Microelectronics Journal}, 45(11):1380 -- 1391, 2014.

\bibitem{KS_ternary}
M.~Warzecha, M.~Oszajca, K.~Pilarczyk, and K.~Szaciłowski.
\newblock A three-valued photoelectrochemical logic device realising accept
  anything and consensus operations.
\newblock {\em Chem. Commun.}, 51:3559--3561, 2015.

\bibitem{whiting2018parallel}
James~GH Whiting, Richard Mayne, and Andrew Adamatzky.
\newblock A parallel modular biomimetic cilia sorting platform.
\newblock {\em Biomimetics}, 3(2):5, 2018.

\bibitem{williams1989learning}
Ronald~J Williams and David Zipser.
\newblock A learning algorithm for continually running fully recurrent neural
  networks.
\newblock {\em Neural computation}, 1(2):270--280, 1989.

\bibitem{yi2016fpga}
Yang Yi, Yongbo Liao, Bin Wang, Xin Fu, Fangyang Shen, Hongyan Hou, and Lingjia
  Liu.
\newblock Fpga based spike-time dependent encoder and reservoir design in
  neuromorphic computing processors.
\newblock {\em Microprocessors and Microsystems}, 46:175--183, 2016.

\bibitem{RN640}
Konkoli Zoran.
\newblock A perspective on {P}utnam’s realizability theorem in the context of
  unconventional computation.
\newblock {\em International Journal of Unconventional Computing}, 11:83--102,
  2015.

\end{thebibliography}

\end{document}